\documentclass[useAMS, usenatbib,usegraphicx]{mn2e}
\usepackage{rotating}

\newcommand{\msunpc}{M$_\odot$ pc$^{-2}$}

\newcommand{\Ttheta}{$\Theta$}
\newcommand{\data}{${\bf \mathcal{D}}$}
\newcommand{\prob}{${\mathcal{P}}$}

\newcommand{\sigscat}{$\sigma_{\rm scat}$}
\newcommand{\Halpha}{H$\alpha$}

\newcommand{\sigmol}{$\Sigma_{\rm mol}$}
\newcommand{\sigsfr}{$\Sigma_{\rm SFR}$}

\newcommand{\ICOmeas}{$\hat{I}_{\rm CO}$}
\newcommand{\sigmolmeas}{${\hat \Sigma_{\rm mol}}$}
\newcommand{\sigsfrmeas}{${\hat \Sigma_{\rm SFR}}$}

\newcommand{\Xunits}{cm$^{-2}$ K$^{-1}$ km$^{-1}$ s}

\newcommand {\apgt} {\ {\raise-.5ex\hbox{$\buildrel>\over\sim$}}\ }
\newcommand {\aplt} {\ {\raise-.5ex\hbox{$\buildrel<\over\sim$}}\ }
\newcommand{\XCO}{$X_{\rm CO}$}

\newcommand{\epsscat}{$\epsilon_{\rm scat}$}
\newcommand{\KS}{Kennicutt-Schmidt}
\newcommand{\N}{$N$}
\newcommand{\A}{$A$}
\newcommand{\chisq}{$\chi^2$}
\newcommand{\OLSyx}{OLS$(\Sigma_{\rm SFR}|\Sigma_{\rm mol})$}
\newcommand{\OLSxy}{OLS$(\Sigma_{\rm mol}|\Sigma_{\rm SFR})$}
\newcommand{\tsig}{$2\sigma$}
\newcommand{\osig}{$1\sigma$}

\newcommand{\ppm}{$\pm$}

\title[A non-universal molecular KS relationship] {Evidence for a non-universal Kennicutt-Schmidt relationship using hierarchical Bayesian linear regression}
\author[R. Shetty et al.]{Rahul Shetty$^{1}$, Brandon C. Kelly$^{2}$, Frank Bigiel$^{1}$ \\
$^{1}$ Zentrum f\"ur Astronomie der Universit\"at Heidelberg, Institut f\"ur Theoretische Astrophysik, Albert-Ueberle-Str. 2, 69120 Heidelberg, Germany \\
$^{2}$ Department of Physics, Broida Hall, University of California,
  Santa Barbara, CA 93106, USA}

\begin{document}

\date{Accepted 2012 December 12. Received 2012 December 11; in original form 2012 October 3}

\pagerange{\pageref{firstpage}--\pageref{lastpage}} \pubyear{2012}
\maketitle

\label{firstpage}
\begin{abstract}

For investigating the relationship between the star formation rate and
gas surface density, we develop a Bayesian linear regression method
that rigorously treats measurement uncertainties and accounts for
hierarchical data structure.  The hierarchical Bayesian method
simultaneously estimates the intercept, slope, and scatter about the
regression line of each individual subject (e.g. a galaxy) and the
population (e.g. an ensemble of galaxies).  Using synthetic datasets,
we demonstrate that the method accurately recovers the underlying
parameters of both the individuals and the population, especially when
compared to commonly employed ordinary least squares techniques, such
as the bisector fit.  We apply the hierarchical Bayesian method to
estimate the Kennicutt-Schmidt (KS) parameters of a sample of spiral
galaxies compiled by \citet{Bigiel+08}.  We find significant variation
in the KS parameters, indicating that no single KS relationship holds
for all galaxies.  This suggests that the relationship between
molecular gas and star formation differs from galaxy to galaxy,
possibly due to the influence of other physical properties within a
given galaxy, such as metallicity, molecular gas fraction, stellar
mass, and/or magnetic fields.  In four of the seven galaxies the slope
estimates are sub-linear, especially for M51, where unity is excluded
at the \tsig\ level.  We estimate the mean index of the KS
relationship for the population to be 0.84, with \tsig\ range [0.63,
1.0].  For the galaxies with sub-linear KS relationships, a possible
interpretation is that CO emission is tracing some molecular gas that
is not directly associated with star formation.  Equivalently, a
sub-linear KS relationship may be indicative of an increasing gas
depletion time at higher surface densities, as traced by CO emission.
The hierarchical Bayesian method can account for all sources of
uncertainties, including variations in the conversion of observed
intensities to star formation rates and gas surface densities
(e.g. the \XCO\ factor), and is therefore well suited for a thorough
statistical analysis of the KS relationship.

\end{abstract}


\section{Introduction} \label{introsec}

\subsection{The Kennicutt-Schmidt relationship} \label{introKSsec}

Accurately measuring the properties of star formation is crucial for
understanding a variety of astrophysical topics, including the
interstellar medium (ISM), galaxy structure and dynamics, and the
evolution of the universe as a whole.  Observations have revealed
strong power-law correlations between the star formation rate
\sigsfr\ and gas surface densities in disk galaxies, now referred to
as the Schmidt or Kennicutt-Schmidt (hereafter KS) law
\citep{Schmidt59, Kennicutt89, Kennicutt98, Kennicutt&Evans12}.
Explaining this KS law has been a major priority in the star formation
research community \citep[see e.g.][and references
  therein]{Maclow&Klessen04, McKee&Ostriker07, Kennicutt&Evans12}.

Recent studies have indicated that much of the KS correlation is
driven by the molecular component \sigmol:
\begin{equation}
\Sigma_{\rm SFR} = a \Sigma_{\rm mol}^N,
\label{KSlaw}
\end{equation}
using either azimuthal averages or data from individual sub-kpc
regions \citep{Rownd&Young99, Wong&Blitz02, Heyer+04, Kennicutt+07,
  Bigiel+08, Bigiel+11, Leroy+08, Rahman+11, Schruba+11}.  The
molecular component is usually inferred through CO observations and
traces much of the cold, dense gas in the ISM, because HI saturates
above surface densities $\sim$10 \msunpc.  The range in power-law
indices has been estimated from \aplt0.9 to \apgt3 (see intro in
\citealt{Bigiel+08}, and review by \citealt{Kennicutt&Evans12}), and
depends on a variety of caveats, such as the observed scale, tracer,
calibration of star formation rates \citep[e.g.][]{Calzetti+07,
  Rahman+11, Liu+11, Leroy+12}, and a host of properties of the
source, such as the metallicity, gas fractions, and stellar mass
\citep[e.g.][]{Leroy+08, Shi+11, Saintonge11b}.

Averaged over entire galaxy disks, \citet{Kennicutt89, Kennicutt98}
found that $N$ =1.4 $\pm$ 0.15 describes the KS relationship well for
over five orders of magnitude in total gas surface density \citep[see
also][]{Buat+89}. \citet{Kennicutt+07} and \citet{Liu+11} infer a
similar power-law relationship on $\sim$500-2000 pc scales in the
spiral galaxy M51 (NGC 5194).  An interpretation for $N$=1.5 is that
the dominant timescale is determined by global gravitational
instability \citep{Quirk72, Kennicutt89}.  \citet{Bigiel+08},
hereafter B08, analyzed resolved observations of a sample of galaxies,
and inferred an approximately linear molecular KS relationship at
intermediate densities 10 \msunpc\ $<$ \sigmol\ $<$ 100 \msunpc\, and
speculated about a super-linear relationship at higher densities.  As
discussed by B08 a linear relationship may be evidence of a constant
molecular gas depletion time, and that extragalactic CO observations,
often averaging emission over many square kpc, are simply ``counting''
molecular clouds with relatively similar properties.  For M51, both
B08 and \citet{Blanc+09} find a sub-linear KS relationship.  There are
numerous differences in the observations and analysis, including
reduction techniques, that may account for the discrepancies between
the estimated power-law indices.  Thus, a comparison between the
various observational investigations is often not straightforward.  We
also caution that any analysis and conclusions, including those from
our work here, are affected by the specific assumptions that go into
data reduction or conversions from observables to physical quantities.
However, with these caveats in mind, we will use the B08 data sets to
illustrate the application of a hierarchical Bayesian
method.\footnote{We note that the calibrations for some of these data
  sets may have been revised in the meantime (e.g. Leroy et al 2012,
  submitted), though without changing the overall conclusions
  regarding the KS relationship.  In a future paper we plan to expand
  this analysis to include these most recent calibrations and larger
  galaxy samples.}

\subsection{Linear regression in astrophysics} \label{introlinsec}

A central component of many observational investigations is the
quantification of the correlation between two or more observed
quantities.  Typically, linear regression provides estimates of the
zero-point and slope of the ``best-fit'' regression line between the
observed data.  In log-space, linear regression of the logarithm of
the observed quantities provide estimates of the coefficient $C$ and
index $N$ of a power-law $y=Cx^N$.  For example, linear regression is
often employed to quantify the relationship between the linewidth and
sizes of molecular clouds, the luminosity and kinematic velocity of
galaxies (Tully-Fisher relationship), the X-ray spectral slope and
Eddington ratios of quasars, the star formation rate and gas surface
density in the ISM, and a host of other topics in current
astrophysical research.  It is thus important to understand the
limitations of common fitting methods, and, whenever possible, develop
accurate regression algorithms appropriate for the given problem.

One common statistical method for fitting data is the ordinary least
squares (OLS), or \chisq\ fit.  An OLS fit computes the best fit line
by minimizing the squared error in the residuals.  As \citet{Isobe+90}
show, the OLS fit can produce discrepant slope and intercept estimates
depending on the classification of the ``dependent'' and
``independent'' variables.  An important limitation of the OLS method
is that it does not account for measurement uncertainties in the
regression, resulting in biased parameter estimates
\citep{Akritas&Bershady96, Weiner+06, Kelly07}.

Further, observational datasets are often intrinsically hierarchical,
or structured, but most common statistical methods do not account for
any such hierarchy.  For example, consider \sigsfr\ and \sigmol\ in a
sample of observed galaxies.  A given galaxy within the survey sample
will contain a number of measurements of \sigsfr\ and \sigmol.  If a
linear regression is employed on all measurements of \sigsfr\ and
\sigmol\ from all galaxies, then any galaxy-to-galaxy variation could
be lost in the global \sigsfr-\sigmol\ relationship.  Moreover, the
global or universal relationship estimated from a linear regression
analysis on all data may be biased towards the trend from one or more
galaxies, e.g. those with the largest number of \sigsfr $-$
\sigmol\ pairs, or, if noise is considered, to those galaxies with the
highest signal-to-noise ratios and with the tightest \sigsfr $-$
\sigmol\ correlations.  Here, we develop a fitting method for such
hierarchical data, such that we estimate regression parameters for
both the population, or group, as well as the individual relationship,
allowing for an assessment of the differences between individuals
within the group.

In this work, we present a method which rigorously treats measurement
uncertainties, as well as estimates the scatter about the fit
regression lines, two aspects which can naturally be handled in a
hierarchical framework.  Bayesian methods are well suited for treating
problems with hierarchical structure
\citep[e.g.][]{Gelman+04,Gelman&Hill07,Kruschke11}.  Moreover,
measurement uncertainties can be rigorously treated at each level in
the hierarchy, such that the estimated parameters fully account for
any source of uncertainty in the modelling.  Bayesian methods are
becoming more common in astrophysics \citep{Loredo12}, and have been
employed for a number of problems, such as the correlation between
mass and richness of galaxy clusters \citep{Andreon&Hurn10}, binary
eccentricity \citep{Hogg+10}, supernova light curves
\citep{Mandel+11}, modelling dust SEDs \citep{Kelly+12} and extinction
laws \citep{Foster+12}, and turbulence in the ISM \citep{Shetty+12},
to name a few applications.  Here, we introduce a general hierarchical
Bayesian method for linear regression, and to illustrate its
applicability we estimate the parameters of the KS relationship in
local galaxies.

In the next Section, we provide an overview of hierarchical modelling,
and describe the hierarchical Bayesian framework.  We demonstrate the
accuracy of the method on synthetic datasets in Section \ref{testsec}.
After a brief discussion on the observational datasets in Section
\ref{obssec}, we present results from the application of the method on
the B08 observations in Section \ref{ressec}.  In Section
\ref{summarysec}, after listing some caveats and future prospects, we
offer an interpretation of our results and conclude with a summary.

\section{Modelling Method} \label{methosec}

In this section, we describe the modelling method we employ to
estimate the parameters of the KS relationship.  As Bayesian inference
is still not extensively employed in astrophysics research, we
describe why it may be preferred over traditional methods.  We begin
by motivating hierarchical modelling, in the context of the KS law.
We then provide a general description of Bayesian inference, followed
by the modelling method and an outline of the fitting routine.
A general introduction to Bayesian analysis can be found in
\citet{Kruschke11}.  For thorough descriptions of hierarchical
Bayesian methods, we refer the reader to \citet{Gelman+04} and
\citet{Gelman&Hill07}.

\subsection{Motivation: hierarchical modelling with measurement error} \label{motivsec}

The parameters of the Kennicutt-Schmidt (KS) relationship are governed
by Equation \ref{KSlaw}, relating \sigsfr\ and \sigmol.  One of our
primary goals is to estimate the distribution of KS parameters for a
population of galaxies, e.g. the mean and dispersion in KS slopes for
the galaxy population.  Estimating the KS relationship requires
measuring \sigsfr-\sigmol\ pairs within a galaxy, and for many
individual galaxies.  As such, the process of estimating a universal
KS relationship is intrinsically {\it
  hierarchical}.\footnote{Different terminology, including
  ``multilevel'' or ``random effects'' has been used in place of
  ``hierarchical'' modelling \citep[e.g.][]{Gelman+04,
    Gelman&Hill07}.}

Of course, the star formation properties in each galaxy will likely be
influenced by physical processes that may or may not be directly
associated with the surface density \citep[e.g.][]{Leroy+08}.  Local
or large scale processes, for example, metallicity, magnetic fields,
molecular gas fractions, turbulent levels, or rotation may all
influence the star formation properties, besides \sigmol.  Moreover,
\citet{Shi+11} find that \sigsfr\ portrays a stronger correlation with
the stellar surface density compared to the gas surface density across
galaxies with different morphologies.  Therefore, the \sigsfr -
\sigmol\ relationship in different galaxies may not necessarily be
expected to follow a single trend, and there may be scatter about any
fit KS law as formulated by Equation \ref{KSlaw}.  Indeed, uniform
analyses of a sample of galaxies by B08 has resulted in a range of
indices 0.84 - 1.12.

Another unavoidable caveat in fitting a model is the effect of
measurement uncertainties.  Measurement uncertainties can produce
significant biases when fitting a model
\citep[e.g.][]{Akritas&Bershady96, Weiner+06,Kelly07}.  These
uncertainties will also contribute to the scatter about any regression
line, which should be quantified during the fit.  Further, there are
additional sources of uncertainty in the conversion between
observables (e.g., UV, IR or CO intensity) and \sigsfr\ and \sigmol\
needed for evaluating the KS relationship.  In particular, the
measurements of \sigmol\ are strongly dependent on the adopted value
of the ``\XCO\ factor'', the conversion of CO intensity to H$_2$
surface density, for which recent efforts have clearly demonstrated
varies with environment \citep[see e.g.][]{Glover&Maclow11,
  Shetty+11a, Shetty+11b, Narayanan+12, Feldmann+12a, Sandstrom+12}.
In this work, we focus on the estimated measurement uncertainties
(e.g. due to calibration) on \sigsfr\ and \sigmol\ directly, though we
preview future efforts where \XCO\ will be treated self-consistently
in the modelling method.

Our goals are to estimate the parameters of the KS relationship for
each individual galaxy as well as the universal values, including a
rigorous treatment of uncertainties and the scatter about the
regression line.  On the individual galaxy level, each galaxy is
considered to have its own relationship between \sigsfr\ and \sigmol.
The parameters governing the \KS\ relationship are the power-law index
\N\ and the coefficient, \A.  After transforming Equation \ref{KSlaw}
to log space, we have:
\begin{equation}
\log(\Sigma_{\rm SFR}) = A + N \log (\Sigma_{\rm mol}) + \epsilon_{\rm scat}
\label{KSlaw_log}
\end{equation}
where $A=\log(a)$.  The scatter about the regression line is \epsscat,
assumed to have mean 0 and dispersion \sigscat, which is one of the
parameters to be estimated from the method.

In the evaluation of Equation \ref{KSlaw_log}, the parameters for each
galaxy should be related to the universal values of those parameters.
Under a hierarchical framework, both the individual and universal, or
group, parameters are estimated simultaneously.  We carry out the fit
under a Bayesian framework, which is ideally suited for evaluating all
the parameters of a hierarchical model.



\subsection{Bayesian Inference}

Bayes' Theorem allows for the evaluation of probability \prob\ of a
set of parameters \Ttheta\ given the observed data \data:
\begin{equation}
\mathcal{P}({\Theta | \mathcal{D}})  \propto \mathcal{P}({\mathcal{D}|\Theta})\mathcal{P}(\Theta).
\label{bayest}
\end{equation}
Here, $\mathcal{P}(\Theta)$ is the prior on \Ttheta, where \Ttheta\ is
a vector of all the parameters defining a model.  The parameters
making up the set \Ttheta\ are described below, and includes the slope
of the KS law $N$, for each individual as well as the group value.
The other term on the right hand side,
$\mathcal{P}({\mathcal{D}|\Theta})$, is the likelihood, which is the
probability of the data given the set \Ttheta.  The outcome of
Bayesian inference is the posterior $\mathcal{P}({\Theta |
  \mathcal{D}})$, which is the probability distribution function (PDF)
of the model parameters \Ttheta\, given the data \data.

The next two subsections describe the measurement and the full
hierarchical model, defining the likelihood and priors.  We use
standard statistical notation in describing how quantities are
conditionally related and their distributions.  Namely, $y|x$
indicates a variable $y$ given a value of $x$.  And, $y | \mu,
\sigma^2 \sim \mathcal{N}(\mu,\sigma^2)$ denotes that $y$ is drawn
from a normal distribution $\mathcal{N}$, given a mean value $\mu$ and
variance $\sigma^2$.  The mean value of a vector $x$ is denoted by
$\overline{x}$.  In the model, we use Gamma functions $\Gamma(s,r)$
for the distributions on the inverse of the variance, with $s$ and $r$
the shape and rate parameters.\footnote{The inverse of the variance is
  referred to as the precision, for which $\Gamma$ distributions are
  commonly employed for their distributions \citep{Kruschke11,
    Gelman+04}.}

\subsection{The measurement model} \label{uncsec}

Due to observational uncertainties, the measured values
\sigmolmeas\ and \sigsfrmeas\ are related to the true values
\sigmol\ and \sigsfr\ by

\begin{eqnarray}
\log \hat{\Sigma}_{\rm SFR} = \log \Sigma_{\rm SFR} + \epsilon_{\rm
  SFR} \label{measmod1} \\
\log \hat{\Sigma}_{\rm mol} = \log \Sigma_{\rm mol} + \epsilon_{\rm
  mol} \label{measmod2}
\end{eqnarray}
Here, $\epsilon_{\rm SFR}$ and $\epsilon_{\rm mol}$ are the random
measurement errors, assumed to have mean 0 and fixed known (or
estimated) dispersion $\sigma_{\rm SFR}$ and $\sigma_{\rm mol}$,
respectively.

\subsection{The Hierarchical Model} \label{hiermsec}

For each galaxy $j$ ($j=1, 2, ... , J$), observations provide
measurements ${\hat \Sigma}_{\rm mol, ij}$, ${\hat \Sigma}_{\rm SFR,
  ij}$, where $i$ ($i=1, 2, ... , I_J$) indicates individual
measurements within a given galaxy $J$.  Along with those
measurements, we have their estimated uncertainties parametrised by
$\sigma_{\rm mol, ij}$ and $\sigma_{\rm SFR, ij}$.  As the desired KS
parameters relate the ``true'' values of \sigsfr\ with \sigmol, we
need to estimate \sigsfr\ and \sigmol\ from the measurements and
uncertainties.  The following set of conditional probability
distributions denotes the relationships between those parameters on
the individual galaxy level in the hierarchy, as well as the
relationship between the KS parameters and the true values of
\sigmol\ and \sigsfr:
\begin{eqnarray}
\log\hat{\Sigma}_{\rm mol, ij} | \Sigma_{\rm mol, ij} \sim \mathcal{N}(\log \Sigma_{\rm mol, ij} , \sigma^2_{\rm mol, ij}) \label{indh1}  \\
\log\Sigma_{\rm mol, ij} | \overline{\Sigma}_{\rm mol, j}, \overline{\sigma^2}_{\rm mol, j} \sim \mathcal{N}(\log \overline{\Sigma}_{\rm mol, j}, \overline{\sigma^2}_{\rm mol, j} ) \label{indh2} \\
\log\hat{\Sigma}_{\rm SFR, ij} | \Sigma_{\rm SFR, ij} \sim \mathcal{N}(\log \Sigma_{\rm SFR, ij} , \sigma^2_{\rm SFR, ij}) \label{indh3} \\
\log \Sigma_{\rm SFR, ij} | A_j, N_j, \Sigma_{\rm mol,ij}, \sigma^2_{\rm scat, j}  \sim \label{indh4} \\ \nonumber
\mathcal{N}(A_j + N_j \log \Sigma_{\rm mol,ij} , \sigma^2_{\rm scat, j})
\label{KSinhier}
\end{eqnarray}
We have constructed a model using normal distributions for all the
relevant parameters in the individual galaxy level.  Note that
Equation \ref{indh4} contains the KS relationship from Equation
\ref{KSlaw_log}.  The relationships above require quantities that must
be evaluated in the group level of the hierarchy, i.e., those related
to the distribution of KS parameters for the galaxy population, such
as $A_j, N_j, \sigma_{\rm scat, j}$.

For the group model, we have:
\begin{eqnarray}
A_j \,|\, \mu_{\rm A}, v_{\rm A} \sim \mathcal{N}(\mu_{\rm A}, v_{\rm A})  \label{grprior1} \\
N_j \,|\, \mu_{N}, v_{N} \sim \mathcal{N}(\mu_{N}, v_{N}) \label{grprior1b} \\
\log \overline{\Sigma}_{\rm mol, j} \, | \, \mu_{\rm mol}, v_{\rm mol} \sim \mathcal{N}(\mu_{\rm mol}, v_{\rm mol}) \\
1/\overline{\sigma^2}_{\rm mol, j} \,|\, s_{\rm mol}, r_{\rm mol} \sim \Gamma(s_{\rm mol}, r_{\rm mol}) \\
1/\sigma^2_{\rm scat, j} \,|\, s_{\rm scat}, r_{\rm scat} \sim \Gamma(s_{\rm scat}, r_{\rm scat})
\label{grprior2}
\end{eqnarray}
Equations \ref{indh1} $-$ \ref{grprior2} describe the quantities for
the individuals and the population.  Those quantities with subscript
$j$ refer to individual galaxy properties, for instance
$\overline{\Sigma}_{\rm mol, j}$ is the mean gas surface density of
galaxy $j$, and $\mu_{\rm mol}$ is the mean value of the surface
density for all galaxies.  These equations describe how the
distributions of individual parameters are derived from the group
parameters.  For example, the individual slopes $N_j$ are conditional
on the group slopes and variances $\mu_N$ and $v_N$, respectively.
Similarly, the distributions of the group values require assumed
distributions on another level in the hierarchy.

The final hierarchical level completes the model setup.  The assumed
distributions of quantities in this level are the ``hyperpriors''
which govern the group distributions, which in turn lead to the
individual parameters.  We construct broad hyperpriors, as we would
like to rely primarily on the data to estimate the group parameters:

\begin{eqnarray}
\mu_{\rm A} \sim \mathcal{N}(0, 100) \label{indhp1a} \\
1/v_{\rm A} \sim \Gamma(0.1, 0.1) \\
\mu_{\rm N} \sim \mathcal{N}(0, 100) \label{indhp1b} \\
1/v_{\rm N} \sim \Gamma(0.1, 0.1) \\
\nonumber \\
\mu_{\rm mol} \sim \mathcal{N}(0, 100) \\
1/v_{\rm mol} \sim \Gamma(0.1, 0.1) \\
\nonumber \\
s_{\rm mol} \,|\, m_{\rm mol}, d_{\rm mol} = m_{\rm mol}^2/d_{\rm mol}^2 \\
r_{\rm mol} \,|\, m_{\rm mol}, d_{\rm mol} = m_{\rm mol}/d_{\rm mol}^2 \\
m_{\rm mol} \sim \Gamma(1, 0.1) \\
d_{\rm mol} \sim \Gamma(1, 0.1) \\  \nonumber
\\
s_{\rm scat} \,|\, m_{\rm scat}, d_{\rm scat} = m_{\rm scat}^2/ d_{\rm scat}^2 \\
r_{\rm scat} \,|\, m_{\rm scat}, d_{\rm scat} = m_{\rm scat}/d_{\rm scat}^2 \\
m_{\rm scat} \sim \Gamma(1, 0.1) \\
d_{\rm scat} \sim \Gamma(1, 0.1)  \label{indhp2}
\end{eqnarray}

The hierarchical model described in Equations \ref{indh1} -
\ref{indhp2} contains all the relevant conditional dependencies and
distributions to evaluate Equation \ref{bayest}.  The hyperpriors of
the group KS parameters (Eqns. \ref{indhp1a} and \ref{indhp1b}) are
normally distributed, with very large variances.  These wide
distributions allow the data to govern the final estimates of the PDFs
of $\mu_A$ and $\mu_N$.  For the variances of the individual
parameters, $v_A$ and $v_N$, as well as for the scatter term, their
inverses are modeled as Gamma functions, for which the estimated
values are again primarily governed by the data.

The main parameters we are interested in are those that define the KS
relationship, at both the individual and group levels, and the scatter
term about the regression line in Equation \ref{KSlaw_log}:
$\Theta^\prime = (A_j, N_j, \mu_A, \mu_N, \sigma^2_{\rm scat, j})$.
We simply marginalize over the other parameters required for
estimating the regression parameters.

Modern Markov Chain Monte Carlo (MCMC) techniques allow for efficient
sampling of the full parameter space.  We employ the ``Gibbs
sampling'' method for generating random draws from the posterior.  The
posterior is constructed by multiplying the conditional relationships
defined by Equations \ref{indh1} - \ref{indhp2}.  At each step in the
MCMC chain, Gibbs sampling generates a random draw from the posterior
by cycling through the conditional distributions of each parameter,
such that a new value of each parameter is drawn from its distribution
conditional on the current values of all the other parameters and the
data \citep[see, e.g.,][]{Gelman+04}.  We use the {\tt JAGS} software
\citep[Just Another Gibbs Sampler,][]{JAGS}\footnote{{\tt JAGS} is
freely available from http://mcmc-jags.sourceforge.net.} within the
{\tt R} programming language\footnote{{\tt R} is freely available from
http://cran.r-project.org.} to carry out this analysis.  In our
execution, we run three seperate MCMC chains, with each chain
containing 25,000 steps.  With this choice of a large number of
sampling steps, the convergence of parameter estimates is easily
achieved.  Before proceeding to evaluate this model on the observed
data, we use synthetic data with known parameters to verify the
accuracy of this hierarchical framework.

\section{Method Testing} \label{testsec}

To test the accuracy of the method, we apply it to synthetic datasets,
and compare the parameter estimates to the adopted values.  As we
describe below, many aspects of the model assumptions are not strictly
satisfied.  These discrepancies allow us to test the sensitivity of
the model assumptions on the derived parameter estimates.

For our initial tests, the data are constructed to resemble the
observed sample we analyze in Section 5.  We construct two groups,
each containing 7 individual galaxies, which is the number of galaxies
in the B08 sample we analyze in the next Section.  For each galaxy, we
choose values for the slope, intercept, and scatter and evaluate
Equation \ref{KSlaw_log}.  The chosen slopes and intercepts are
results from fitting the B08 sample: for Group A, the Bayesian
inferred values from this work (Section \ref{ressec}), and for Group
B, the bisector fits (from B08).  We generate 50 log(\sigmol) values
for each galaxy, which are uniformly distributed between 0.1 and 2.2,
a range comparable to log(\sigmol) from B08.  We add a value drawn
from a Gaussian distribution with 0 mean and 0.1 standard deviation
for the intrinsic scatter term in Equation \ref{KSlaw_log}.  For the
50 simulated values of \sigmol\ and \sigsfr\ of each individual
galaxy, we construct \sigmolmeas\ and \sigsfrmeas\ by adding noise
drawn from normal distributions with zero mean and standard deviations
corresponding to 25\% and 50\% of \sigmol\ and \sigsfr.  These noise
levels are the estimates provided in \citet{Leroy+09} and
\citet{Bigiel+10c}.  We then fit the hierarchical model over the full
range of \sigsfrmeas\ and \sigmolmeas.  The first four columns of
Tables \ref{bayesgrpA} and \ref{bayesgrpB} show the intercepts,
slopes, and scatter terms of the test galaxies in groups A and B,
respectively.  The group parameters in the last row are simply the
mean values of the intercept and slopes of the individual galaxies.


\subsection{Bayesian Parameter Estimates} \label{bayestestsec}

The results of the Bayesian fit on the datasets are shown in Columns 5
$-$ 5 in Tables \ref{bayesgrpA} and \ref{bayesgrpB}.  The first four
columns correspond to the individual test galaxies, indicating the
adopted values of the intercepts, slopes, and scatter terms,
respectively.  The fifth and sixth column show the median and \tsig\
(95\%) range in the estimated intercepts of each individual test
galaxy.\footnote{It is common practice to conservatively consider the
  95\% interval as the range in plausible parameter estimates
  \citep[see, e.g.][]{Gelman+04, Kruschke11}.}  Similarly, the seventh
and eighth column shows the estimated slopes, and its \tsig\ range.
The last column shows the posterior median of the scatter term
$\sigma_{\rm scat}$ in Equation \ref{KSlaw_log} .  The last row of the
tables shows the adopted and estimated quantities of the group
distribution.

\begin{table*}
 \centering
 \begin{minipage}{140mm}
  \caption{Adopted and Bayesian inferred parameters for Test Group A}
  \begin{tabular}{ccccccccc}
  \hline
  \hline
 Subject & True $A$ & True $N$ & True \sigscat & Bayes $A$ & Bayes $2\sigma_A$ & Bayes $N$ & Bayes $2\sigma_N$ & Bayes \sigscat  \\
\hline
Test Galaxy A1  & $-$2.77 & 0.72 & 0.1 & $-$2.79 & [$-$2.9, $-$2.7] & 0.74 & [0.64, 0.84] & 0.12 \\
Test Galaxy A2  & $-$3.21 & 0.88 & 0.1 & $-$3.23 & [$-$3.4, $-$3.1] & 0.86 & [0.76, 0.97] & 0.12 \\
Test Galaxy A3  & $-$3.18 & 0.89 & 0.1 & $-$3.14 & [$-$3.3, $-$3.0] & 0.89 & [0.79, 0.99] & 0.12 \\
Test Galaxy A4  & $-$2.81 & 0.78 & 0.1 & $-$2.91 & [$-$3.0, $-$2.8] & 0.82 & [0.72, 0.92] & 0.12 \\
Test Galaxy A5  & $-$2.87 & 0.74 & 0.1 & $-$2.91 & [$-$3.0, $-$2.8] & 0.76 & [0.66, 0.86] & 0.12 \\
Test Galaxy A6  & $-$3.22 & 0.91 & 0.1 & $-$3.13 & [$-$3.3, $-$3.0] & 0.87 & [0.78, 0.98] & 0.12 \\
Test Galaxy A7  & $-$2.82 & 0.92 & 0.1 & $-$2.78 & [$-$2.9, $-$2.7] & 0.90 & [0.80, 1.00] & 0.12 \\
\hline
{\bf Group A Parameters$^1$}  & {\bf$-$2.98} & {\bf 0.83} & 0.1 & {\bf $-$2.99} & {\bf [$-$3.2, $-$2.7]} & {\bf 0.84} & {\bf [0.65, 1.0]} & 0.12 \\
\hline
\footnotetext[0]{$^1$ $2\sigma$ range in estimated dispersions of the group intercept and slope are $\sqrt{v_{\rm A}}=$[0.17, 0.57], and $\sqrt{v_N}=$[0.13, 0.42].}
\end{tabular}
\label{bayesgrpA}
\end{minipage}
\end{table*}

\begin{table*}
 \centering
 \begin{minipage}{140mm}
  \caption{Adopted and Bayesian inferred parameters for Test Group B}
  \begin{tabular}{ccccccccc}
  \hline
  \hline
 Subject & True $A$ & True $N$ & True \sigscat & Bayes $A$ & Bayes $2\sigma_A$ & Bayes $N$ & Bayes $2\sigma_N$ & Bayes \sigscat  \\
\hline
Test Galaxy B1  & $-$2.29 & 0.84 & 0.1 & $-$2.30 & [$-$2.4, $-$2.2] & 0.85 & [0.75, 0.95] & 0.12 \\
Test Galaxy B2  & $-$2.53 & 0.92 & 0.1 & $-$2.56 & [$-$2.7, $-$2.4] & 0.91 & [0.81, 1.01] & 0.12 \\
Test Galaxy B3  & $-$2.15 & 0.96 & 0.1 & $-$2.48 & [$-$2.6, $-$2.3] & 0.96 & [0.86, 1.06] & 0.12 \\
Test Galaxy B4  & $-$2.26 & 0.92 & 0.1 & $-$2.35 & [$-$2.5, $-$2.2] & 0.95 & [0.86, 1.05] & 0.12 \\
Test Galaxy B5  & $-$2.33 & 1.00 & 0.1 & $-$2.36 & [$-$2.5, $-$2.2] & 1.01 & [0.90, 1.11] & 0.12 \\
Test Galaxy B6  & $-$2.54 & 1.12 & 0.1 & $-$2.45 & [$-$2.6, $-$2.3] & 1.08 & [0.97, 1.18] & 0.12 \\
Test Galaxy B7  & $-$2.12 & 0.95 & 0.1 & $-$2.09 & [$-$2.2, $-$2.0] & 0.94 & [0.84, 1.06] & 0.12 \\
\hline
{\bf Group B Parameters$^1$}  & {\bf$-$2.37} & {\bf 0.96} & 0.1 & {\bf $-$2.37} & {\bf [$-$2.6, $-$2.1]} & {\bf 0.96} & {\bf [0.77, 1.14]} & 0.12  \\
\hline
\footnotetext[0]{$^1$ $2\sigma$ range in estimated dispersions of the group intercept and slope are $\sqrt{v_{\rm A}}=$[0.15, 0.53], and $\sqrt{v_N}=$[0.13, 0.43].}
\end{tabular}
\label{bayesgrpB}
\end{minipage}
\end{table*}

The median of the posterior, which provides estimates of the most
probable parameters, is similar to the true values.  Within 95\%, the
posterior distributions of the slope and intercept all contain the
true value.  The median of the scatter term is also an accurate
estimate of \sigscat.\footnote{The 95\% confidence interval for
  \sigscat\ is not shown, but is $\sim$[0.09 $-$ 0.45], and always
  contains the true value (0.1).}  This indicates that the
hierarchical fitting has accurately recovered the true parameters.  It
is also evident that the PDFs of the estimated group parameters
include the true values, demonstrating that the method accurately
recovers both the group and individual parameters, to a high degree of
accuracy.

\begin{figure*}
\includegraphics*[width=140mm]{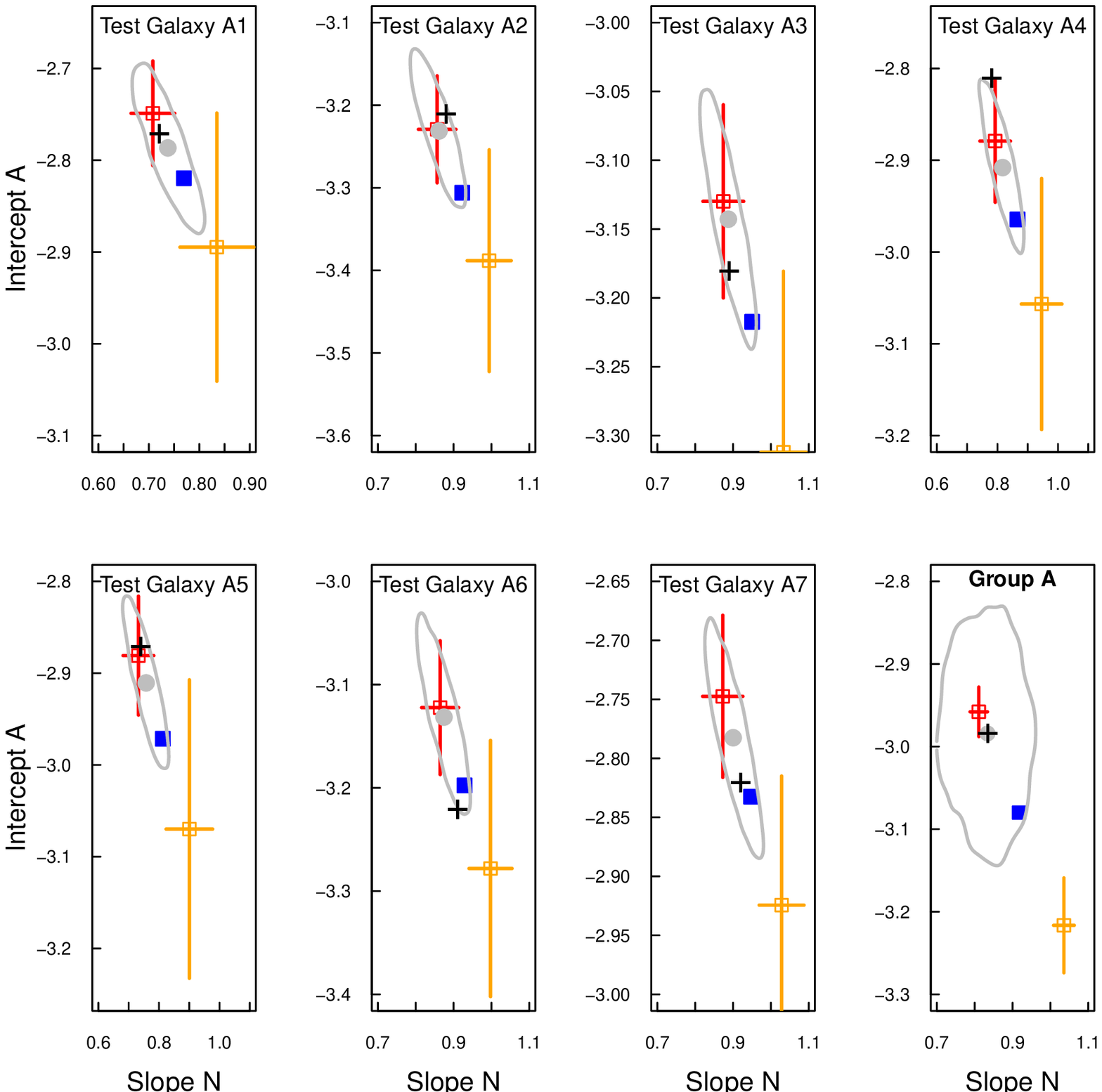}
\caption{ Slope and intercept of test galaxies in Group A.  Black
  cross shows the true values.  Red and orange squares show the
  \OLSyx\ and \OLSxy\ results, with their \osig\ uncertainties,
  respectively.  The gray circles indicate the estimate provided by
  the median of hierarchical Bayesian posterior result, and the
  contours mark the \osig\ deviation.  The filled blue squares mark
  the bisector estimates.  The last panel on the bottom row shows the
  group parameters and fit estimates.}
\label{testresA}
\end{figure*}

\begin{figure*}
\includegraphics*[width=140mm]{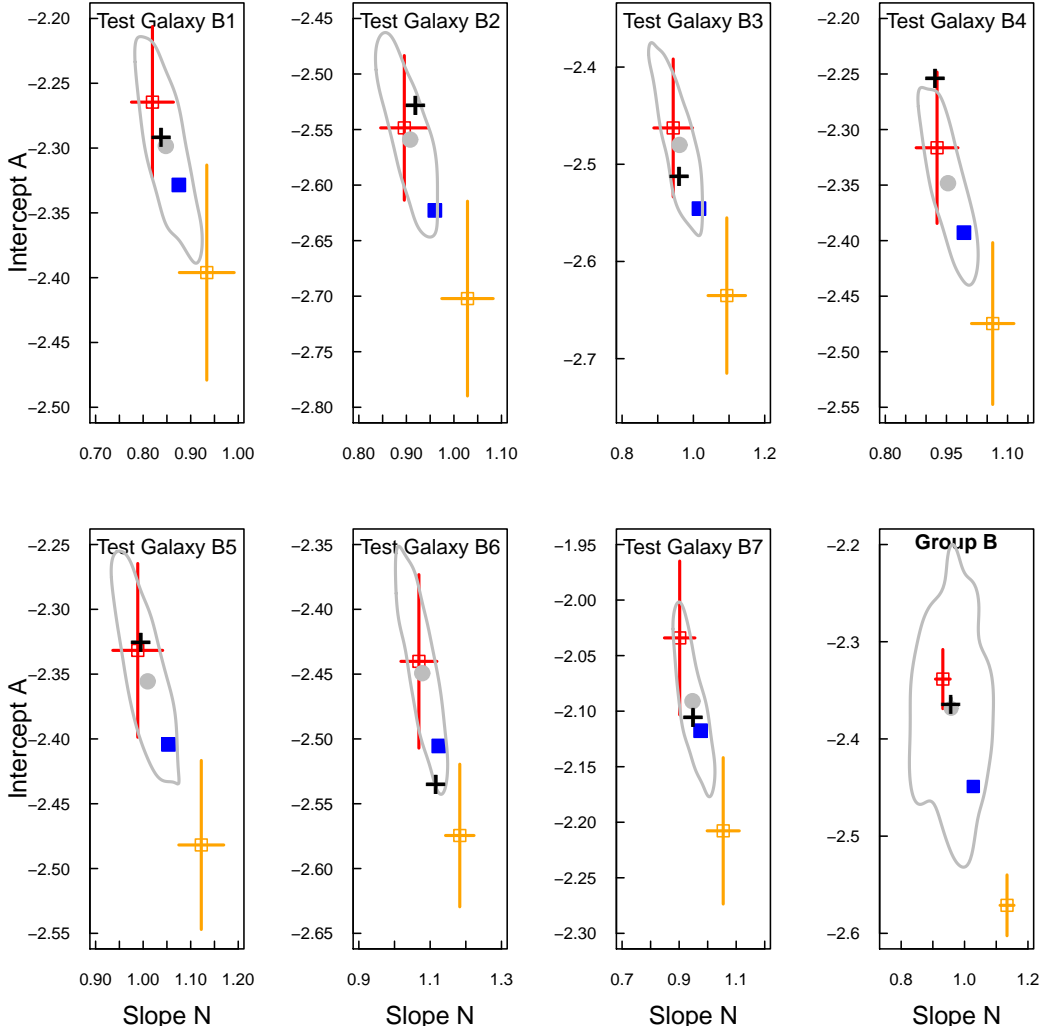}
\caption{ Slope and intercept, along with parameter estimates, of test
  galaxies in Group B.  Symbols and contours are defined in the same
  fashion as in Figure \ref{testresA}.  }
\label{testresB}
\end{figure*}

Figures \ref{testresA} and \ref{testresB} show the median (gray
circle) and \osig\ distribution (gray contour) of the estimated slope
and intercepts, along with the true value (black crosses).  In all but
one case, the \osig\ contours for the individual parameters contains
the true value.  For individual Galaxies A4 and B4, the true value
lies just outside the \osig\ contour, but is within the
\tsig\ interval, as indicated in Tables \ref{bayesgrpA} and
\ref{bayesgrpB}.  In the hierarchical model, the Bayesian estimates of
the individuals are affected by the global group parameters.  This
effect, called ``shrinkage'', drives the overestimate of the slopes
for galaxies A4 and B4 towards the group inferred value, but
nevertheless the true value is contained within the \tsig\ confidence
interval.

\subsection{Comparison with ordinary least squares methods} \label{comptestsec}

In order to assess how the hierarchical Bayesian modelling compares to
other fitting methods, we perform three OLS fits: the $y|x$, $x|y$,
hereafter \OLSyx, and \OLSxy, respectively, and an OLS bisector
\citep[e.g.][]{Isobe+90}.  The estimated slope of the OLS fits is
dependent on the covariance of the data, and variance of the quantity
labeled as the predictor\footnote{The predictor, or covariate, is
  usually the quantity plotted on the ``x''-axis.} or what is
considered the independent quantity, as further discussed below.  In
order not to place a preference on \sigsfr\ or \sigmol, the bisector
fit has emerged as the preferred method in fitting KS relationships.

The estimates from these three OLS fitting methods, however, should
not be interpreted as the same quantities, as discussed by Isobe et
al. (1990, see also \citealt{Gelman&Hill07}).  An OLS($y|x$) estimates
the conditional mean of ``y given x''.  Accordingly, as the
constructed relationships, or simulations, describe how the mean value
of log(\sigsfr) depends on log(\sigmol), the \OLSyx\ is the most
applicable.  The \OLSxy\ estimates how the mean value of log(\sigmol)
depends on log(\sigsfr), and the bisector estimates are weighted means
between the \OLSxy\ and \OLSyx\ - the details of the slope estimates
of the different OLS fits are provided below.  The three OLS results
will provide different estimates, as they are different statistics
derived from the joint distribution of the measurements.  The choice
of which quantity should be identified as the ``x'' or ``y'' variable
should be motivated by the scientific goals.  For the KS law, we are
primarily interested in how \sigsfr\ varies with \sigmol.  As a
result, we would expect the \OLSyx\ to provide more accurate estimates
of the underlying parameters of the simulation for a single galaxy, in
the absence of noise.  Yet, the simulation includes noise and
intrinsic scatter, important properties that a simple \OLSyx\ does not
account for.  As the bisector fit has been employed to estimate the KS
parameters in previous works, we also compare those estimates to the
Bayesian result.  Following conventional practice, for estimating the
group parameters we simply apply the fit to all data from the seven
galaxies together.


Tables 3 and 4 provide the results of the fits.  We show the
\tsig\ uncertainties along with the OLS best fit
parameters.\footnote{For the bisector, the formal uncertainties are
  not given, as they are similar in magnitude to the \OLSyx\ and/or
  \OLSxy\ fit.  A more conservative uncertainty estimate employed by
  B08 is the range in the parameters estimated by \OLSyx\ and \OLSxy.}
These best fit values, along with the \osig\ uncertainties for the
\OLSyx\ and \OLSxy\ estimates, are also shown in Figures
\ref{testresA} and \ref{testresB}.

Tables 3 and 4, as well as Figures \ref{testresA} and \ref{testresB},
indicate that the bisector consistently overestimates the slopes (and
underestimates the intercepts), for all individuals within the groups,
as well as the group estimate.  The results of the \OLSyx\ and \OLSxy\
are rather discrepant, leading to a bisector parameter estimate
falling between those results.

That the estimates from the OLS fitting vary between the three
different methods, as well in comparison to the Bayesian estimates,
can be understood by the definition of the slopes and intercepts in
least-squares fitting.  The OLS parameters have been extensively
discussed and presented \citep[e.g.][]{Isobe+90, Kelly07}, so we
simply state the slope definitions here.  For an \OLSyx\ fit, the
slope $N_{\Sigma_{\rm SFR}|\Sigma_{\rm mol}}$ is computed by taking
the ratio of the covariance (Cov) between the data and the variance
(Var) in the predictor \sigmol:

\begin{equation}
N_{\Sigma_{\rm SFR}|\Sigma_{\rm mol}} = \frac{{\rm Cov} ( {\hat \Sigma_{\rm mol}}, {\hat \Sigma_{\rm SFR})}}{{\rm Var}({\hat \Sigma_{\rm mol}})}
\label{slopeyx}
\end{equation}
For the \OLSxy, the fit slope is the inverse of the desired quantity
in Equation \ref{KSlaw_log}, so that:
\begin{equation}
N_{\Sigma_{\rm mol}|\Sigma_{\rm SFR}} = \frac{{\rm Var}({\hat \Sigma_{\rm SFR})}}{{\rm Cov} ( {\hat \Sigma_{\rm mol}},{\hat \Sigma_{\rm SFR}})}
\label{slopexy}
\end{equation}

The bisector slope $N_{\rm Bis}$ is a weighted mean of the \OLSyx\ and
\OLSxy\ slopes.
\begin{eqnarray}
\label{slopebis}
N_{\rm Bis} = (N_{\Sigma_{\rm mol}|\Sigma_{\rm SFR}} + N_{\Sigma_{\rm SFR}|\Sigma_{\rm mol}})^{-1} \\ \nonumber
\times \biggl[ N_{\Sigma_{\rm SFR}|\Sigma_{\rm mol}}N_{\Sigma_{\rm mol}|\Sigma_{\rm SFR}} - 1 + \\ \nonumber
\sqrt{(1+N^2_{\Sigma_{\rm SFR}|\Sigma_{\rm mol}})(1+N^2_{\Sigma_{\rm mol}|\Sigma_{\rm SFR}})} \biggl]
\end{eqnarray}

Equations \ref{slopeyx} - \ref{slopebis} illustrate what we stated
earlier: that the three different slope estimates are just three
different statistics (or summaries) derived from the same joint
distribution.  Choosing one estimate over the other does not imply
that one quantity ``causes'' the other, as is sometimes claimed to be
implied by the terminology of ``independent'' and ``dependent''
variables.  However, the different slope estimates do differ in
interpretation.  The \OLSyx\ slope describes how the mean value of
\sigsfr\ varies with \sigmol\, while the \OLSxy\ slope describes how
the mean value of \sigmol\ changes with \sigsfr.  Thus, both OLS
slopes are easily interpretable.  In contrast, the bisector slope is a
weighted average of the two OLS slope, and it is not clear how this
should be interpreted.

The OLS slopes are therefore strongly dependent on the statistical
properties of \sigsfrmeas\ and \sigmolmeas.  For the synthetic data of
both groups, Var(\sigmolmeas) = 0.39, pooling all data from each
galaxy together.  For Group A, Var(\sigsfrmeas) = 0.33, and
Cov(\sigmolmeas, \sigsfrmeas) = 0.32.  In Group B, Var(\sigsfrmeas) =
0.41, and Cov(\sigmolmeas, \sigsfrmeas) = 0.36.  The covariances and
Var(\sigsfrmeas) of the two groups are similar, as the adopted slopes
only differ by $\approx$ 10\%.  More importantly, the covariance is
$<$ 1.  As the covariance occurs in the denominator of the \OLSxy\
slope, but in the numerator of the \OLSyx\ slope (and the variances of
the chosen predictors are similar), the resulting $N_{\Sigma_{\rm
    mol}|\Sigma_{\rm SFR}} > N_{\Sigma_{\rm SFR}|\Sigma_{\rm mol}}$.

Notice that the \OLSyx\ parameter estimates are closer to the true
value than the \OLSxy\ estimates.  This is to be expected, because as
described above the \OLSyx\ estimates how the mean value of
log(\sigsfr) depends on log(\sigmol), and the simulated datasets are
constructed with a linear relation between those quantities.  For
these data, the Cov(\sigmolmeas, \sigsfrmeas) is less than the
variance of either quantity, thereby leading to $N_{\Sigma_{\rm
    SFR}|\Sigma_{\rm mol}} < 1$ and $N_{\Sigma_{\rm mol}|\Sigma_{\rm
    SFR}} > 1$.  The \OLSxy\ result therefore drives the bisector
slope towards larger values, leading to the systematic overestimates
shown in Tables 3-4 and Figures \ref{testresA}-\ref{testresB}.  The
overestimate in the bisector slope is not as drastic for Test Group B,
because the underlying slopes are closer to unity to begin with.
Further, note that tests performed by \citet{Isobe+90} (see their
Table 2) produce bisector slopes of unity for a number of scenarios,
where the $y|x$ or $x|y$ slopes are far from unity.  In fact, the
bisector fit is expected to produce a slope of one when $x$ and $y$
are statistically independent, indicative of the difficulty in
interpreting the bisector.  If the scatter about a linear relationship
were very low, for instance due to small measurement uncertainties,
then the \OLSyx\ and \OLSxy, and correspondingly the bisector, would
accurately recover the regression parameters.  Taken together, one can
assess the accuracy of the bisector estimated regression parameters by
inspecting both the OLS$(y|x)$ and OLS$(x|y)$ results.  If they are
highly discrepant, then the bisector result, which will fall between
the OLS$(y|x)$ and OLS$(x|y)$ estimates, should not be considered to
provide accurate parameter estimates of a linear relationship.

\subsection{Effect of Model Assumptions} \label{priortestsec}

For the two synthetic datasets considered so far, the normal
distributions in the intrinsic scatter and measurement uncertainties
matches the prior distributions in the hierarchical model.  In order
to test the sensitivity of these assumptions, we consider an another
synthetic dataset with uniform distributions for the uncertainty and
scatter.  Additionally, compared with the previous tests the synthetic
dataset we consider here has more individual galaxies, with different
numbers of (\sigmolmeas, \sigsfrmeas) pairs for each individual.

The intrinsic slopes and intercepts of each synthetic galaxy in Test
Group C varies between 0.7 to 1.5 and $-$3.0 to $-$2.0, respectively.
The population mean value of the slope is 1.1, and the mean intercept
is $-$2.5.  The intrisic scatter term is uniformly distributed
centered on 0, with extent 0.3.  To construct the noisy measurements
\sigmolmeas\ and \sigsfrmeas, we add a random value drawn from a
uniform distribution centered about 0, with extents 0.2 and 0.25,
respectively.  We carry out the Bayesian fit as before, where we
assume Gaussian distributions, with \osig\ noise estimates equal to
the width of the uniform distributions employed to construct the noisy
datasets.  We also compare the Bayesian results with the direct OLS
fits.

Table 5 shows the intrinsic and Bayesian estimated parameters.  For
the slope and intercept parameter estimates of the individuals, the
true values are always contained within the \tsig\ interval, except
for the slope of Test Galaxy C1 and intercept of Test Galaxy C2, which
have the fewest number of datapoints.  Further, for these individuals
the intrinsic parameters are far from the mean value, so the Bayesian
estimates are affected by shrinkage, which was described above.
Notice that the group parameter estimates also recover the intrinsic
values, and that the posterior median is close to the true values.

The OLS fit results for the population in Table 6 show a marked
difference compared to the Bayesian estimates.  Even at the \tsig\
level, neither the \OLSyx\ and \OLSxy\ can recover the true value.
Consequently, the bisector estimates are also discrepant from the
intrinsic parameters.  For the OLS group estimates, all datapoints
are fit simultaneously, so those individuals with the largest number
of datapoints dominate the final fit.  Therefore, the OLS parameter
estimates tend towards smaller slopes and larger intercepts, thereby
underestimating the group slope and overestimating the intercept, even
when considering the full \tsig\ interval.  This discrepancy,
especially for the \OLSyx\ case, is larger than that from Test Groups
A and B, due to the variable number of datapoints between individuals,
as well as the higher noise and scatter levels.

This test has shown that the Bayesian posterior can reasonably recover
the slopes and intercepts of the individual even though the assumed
distributions for the noise and scatter terms are incorrect.  The
\tsig\ ranges of the posterior only provide erroneous subject
estimates when the individual has very few datapoints, and/or the
individual parameters are far from the population mean values.
Nevertheless, the population values are recovered, indicating a marked
improvement of the Bayesian result compared to the OLS fits.

Notice that the Bayesian estimate of the scatter is systematically
lower than the intrinsic value.  This occurs because the chosen \osig\
value for the noise is set to the extent of the true uniform noise
distribution.  Accordingly, in the synthetic data there are no
measured datapoints occuring at greater than \osig\ from the true
values, whereas the Bayesian model assumes that such noisy measurment
do exist.  Due to this overestimate of the noise properties, the
Bayesian model erroneously designates some of the intrinsic scatter to
be associated with noise uncertainty, and thereby underestimates the
intrinsic scatter.

Clearly, the accuracy of the Bayesian fits are dependent on the
assumed distributions.  If there is good correspondence between the
assumed and intrinsic distributions, as in Test Groups A and B, the
Bayesian results will be highly accurate.  If there is large
discrepancy in the assumed and intrinsic distribution, then the
accuracy Bayesian result will be degraded, as for Test Group C.
Nevertheless, we have shown that for uniformly distributed noise and
scatter properties, the hierarchical Bayesian method employing normal
priors for these properties can nevertheless accurately recover the
population parameters, especially when compared with the OLS methods.
The magnitude of the discrepancy is of course dependent on the noise
and scatter properties.  Further tests may be needed for ascertaining
the reliability of the model on datasets with more discrepant
distributions.  When interpreting the results from the application of
the method to observed datasets, one must of course bear in mind that
the accuracy is dependent on the priors.  As we have considered two
tests in Section \ref{comptestsec} that we consider to be
representative of real observations, we have confidence that the
hierarchical model is appropriate for the observed datasets we analyze
in Section \ref{ressec}.  

\subsection{Summary of method testing} \label{summtestsec}

We have tested the hierarchical Bayesian model described in Section
\ref{methosec} using three sets of synthetic data.  The datasets
consists of groups containing numerous individual galaxies.  Each
galaxy is characterized by some chosen power-law relationship between
\sigsfr\ and \sigmol, as well as intrinsic scatter about this
relationship.  We include noise, similar to estimated uncertainties
from the observations, to produce \sigsfrmeas\ and \sigmolmeas.  The
hierarchical Bayesian method can accurately recover the underlying
parameter estimates, as the peaks of the posterior are very close to
the true values for each individual as well as the mean population
parameters, and the whole \tsig\ range includes the correct solution.

The synthetic datasets do not strictly satisfy all aspects of the
priors in the hierarchical model.  Namely, for Test Groups A and B,
the slope and intercepts of the seven individual galaxies are not
drawn from a normal distribution for the group.  Similarly, the gas
surface densities of each galaxy are not normally distributed, but
rather uniformly distributed.  That the method is able to accurately
recover the underlying slopes and intercepts indicates the
insensitivity of the parameter estimates to these priors.  With Test
Group C, we examine the situation where the intrinsic and noise
distributions are uniformly distributed, which is discrepant from the
assumption of Gaussian distributions in the hierarchical model.  This
group has twenty individuals, each containing a different number of
datapoints.  For this test, the Bayesian fit again accurately recovers
the population slopes and intercepts.  The \tsig\ range of the
posterior only misses the slope and intercept for the individuals that
have the fewest datapoints and the most discrepant slope and intercept
relative to the population values.

The Bayesian fit proves to be more accurate than the OLS fits,
especially compared to the often employed bisector fit.  As discussed,
however, the bisector is expected to estimate a slope which falls
between \OLSyx\ and \OLSxy\ slopes, and is therefore difficult to
interpret.  The bisector should not be considered to provide an
estimate of $N$ in Equation \ref{KSlaw_log}.  The hierarchical
Bayesian result does provide an estimate of this quantity, and its
results are easily interpretable: we can estimate the mean value of
\sigsfr\ given \sigmol, for each individual galaxy as well as for the
population.

\section{Observations} \label{obssec}

We are now in a position to apply the Bayesian method on observational
datasets.  For its first application, the dataset consists of a sample
of seven spiral galaxies from B08, and is publicly available in
  \citet{Bigiel+10c}.  For six of the galaxies, \sigmolmeas\ is
measured in 750 pc-sized regions from the $^{12}$CO $J=2-1$
observations in the HERACLES survey \citep{Leroy+09}.  For the
remaining galaxy, M51, \sigmolmeas\ is computed from $^{12}$CO $J=1-0$
observations from the BIMA SONG survey \citep{Helfer+03}.  The CO
intensities are converted to gas surface densities using a constant
\XCO\ factor, and a constant $^{12}$CO ($J=2-1)/(J=1-0)$ ratio (see
B08 and \citealt{Bigiel+10c} for details) .  We explore variations in
\XCO\ in Section \ref{Xsec}.  To estimate \sigsfrmeas, B08 use the
combination of 24 \micron\ intensities from the SINGS survey
\citep{Kennicutt+03}, and UV fluxes from the GALEX survey
\citep{GildePaz+07}.  We refer the reader to B08 and \citet{Leroy+08,
  Leroy+09} for details of these observations.

\section{Results} \label{ressec}

\subsection{Application to B08 Data} \label{B08res}

The results of the hierarchical Bayesian fit on the seven spiral
galaxies from B08 are shown in Table \ref{bayesgals}.  The median
value of the slopes are all lower than unity, and for four of the
galaxies, NGC 5194 (M51), NGC 5055, NGC 6946, and NGC 628, a linear
slope can be excluded at $>95\%$.  The median of the inferred group
slope is 0.84, with unity falling just inside the \tsig\ interval.

\begin{table*}
\setcounter{table}{6}
 \centering
 \begin{minipage}{140mm}
  \caption{Bayesian estimated parameters for the seven spiral galaxies
    in B08 }
  \begin{tabular}{cccccc}
  \hline
  \hline
 Subject & Bayes $A$ & Bayes $2\sigma_A$ & Bayes $N$ & Bayes $2\sigma_N$ & Bayes \sigscat  \\
\hline
NGC 5194 (M51)   & $-$2.84 & [$-$3.0, $-$2.7] & 0.72 & [0.62, 0.83] & 0.06 \\
NGC 5055   & $-$3.20 & [$-$3.3, $-$3.1] & 0.87 & [0.79, 0.95] & 0.04 \\
NGC 3521   & $-$3.20 & [$-$3.4, $-$3.0] & 0.90 & [0.76, 1.03] & 0.05 \\
NGC 6946   & $-$2.81 & [$-$2.9, $-$2.7] & 0.78 & [0.70, 0.86] & 0.11 \\
NGC 628    & $-$2.89 & [$-$3.1, $-$2.6] & 0.76 & [0.51, 0.95] & 0.05 \\
NGC 3184   & $-$3.24 & [$-$3.4, $-$3.1] & 0.92 & [0.79, 1.10] & 0.05 \\
NGC 4736   & $-$2.83 & [$-$3.2, $-$2.4] & 0.92 & [0.67, 1.20] & 0.08 \\
\hline
{\bf Group Parameters}  & {\bf $-$3.00} & {\bf [$-$3.3, $-$2.7]} & {\bf 0.84} & {\bf [0.63, 1.0]} & 0.14 \\
\hline

\end{tabular}
\label{bayesgals}
\end{minipage}
\end{table*}

\begin{figure*}
\includegraphics*[width=140mm]{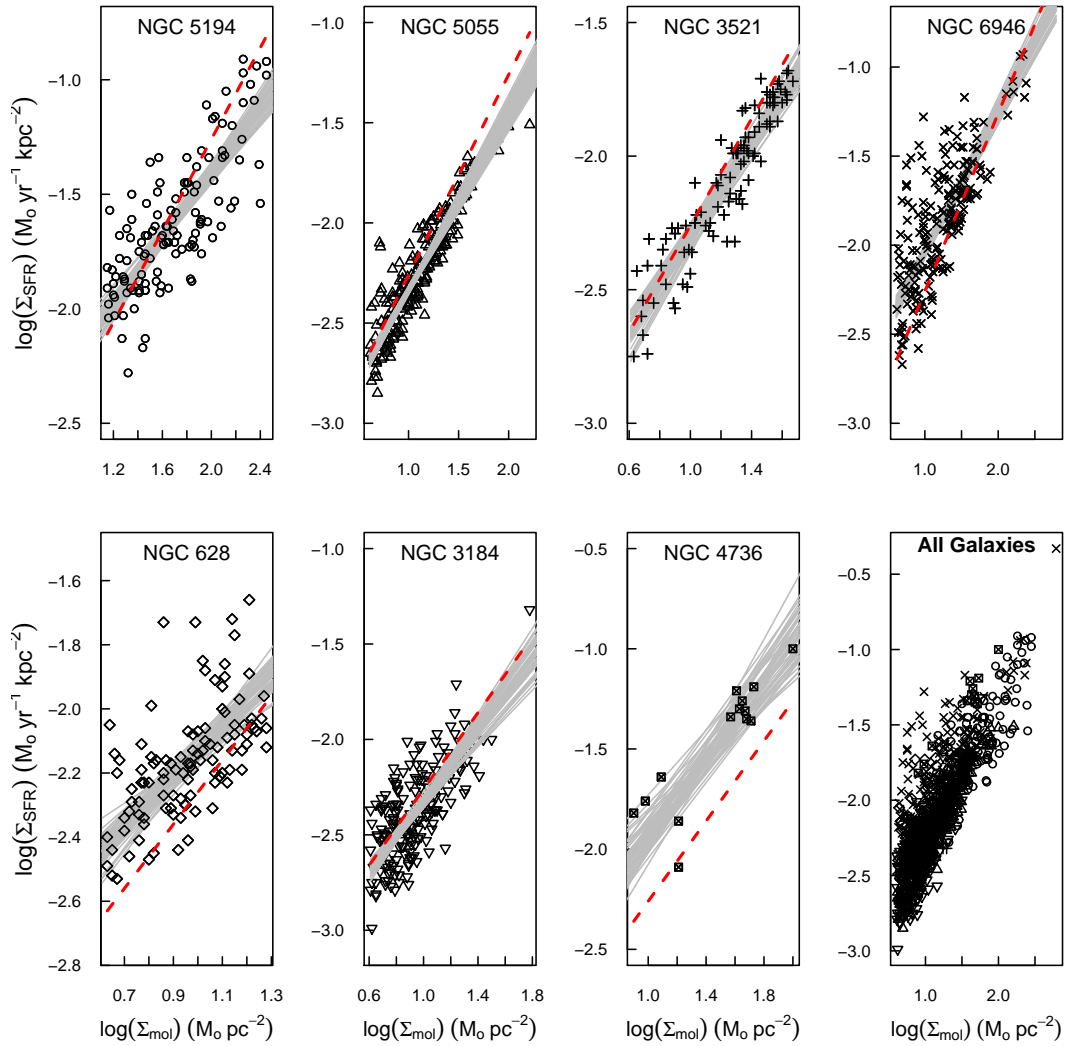}
\caption{Hierarchical Bayesian fits (gray lines), along with the data
  (black symbols), from the B08 sample of seven galaxies (see Fig. 4
  in B08).  Gray lines show 50 random draws from the posterior.  For
  reference, (red) dashed lines show the same linear relationship from
  a bisector fit to all data in the last panel.  Note that the axes
  are different in each panel.}
\label{bayesgalfig}
\end{figure*}

We can verify that the model can reasonably reproduce the data by
over-plotting representative regression lines constructed from the
posterior parameters estimates.  From the posterior PDF, we can draw a
large number of $A$, $N$, and \sigscat\ values, and evaluate Equation
\ref{KSlaw_log} at various \sigmol\ (for each galaxy).  Figure
\ref{bayesgalfig} shows the data and 50 random draws from the
posterior (gray lines) for each galaxy, demonstrating that the
Bayesian model is consistent with the data.  For comparison, the red
dashed line shows the linear relationship inferred from a bisector fit
on all data.  The last panel in Figure \ref{bayesgalfig} only shows
the data from all the galaxies and the associated bisector result.  As
this ensemble is not directly used to infer the group parameters, we
do not overlay the inferred group lines.

The Bayesian inferred parameters are rather different from the
parameters inferred from the bisector fit.  The bisector results are
adopted as the true values of our synthetic dataset in Group B, and
are shown in Table \ref{bayesgrpB} in Section \ref{testsec}.  Recall
that the hierarchical Bayesian fit on that Group accurately recovered
the model parameters, as shown in Figure \ref{testresB} and Table
\ref{bayesgrpB}.  However, the bisector fit produces slope estimates
which are larger than the model slope, and underestimated the
intercept, similar to the results from Group A.  In Group A, the
results of the Bayesian fit from the observed data (Table
\ref{bayesgrpB}) were adopted as the parameters.  For that test group,
the Bayesian result again accurately recovered the adopted parameters.
Recall that the bisector slope is a weighted mean of the \OLSyx\ and
\OLSxy\ estimates.  For the ensemble B08 data, the \OLSyx\ and \OLSxy\
slopes (and \tsig\ uncertainty) are 0.84$\pm$0.04 and 1.19$\pm$0.04,
respectively.  As discussed in Section \ref{comptestsec}, the large
discrepancy between the OLS slope estimates is a clear indicator that
measurement uncertainties in \sigsfr\ and \sigmol\ are affecting the
OLS estimates.

The observed and synthetic data also differ in the detailed statistics
of \sigsfrmeas\ and \sigmolmeas.  For all data pooled together,
Var(\sigmolmeas) = 0.15, Var(\sigsfrmeas) = 0.15, and Cov(\sigmolmeas,
\sigsfrmeas) = 0.12.  These values are a factor of 2$-$3 lower than
the statistics of the synthetic data discussed in Section
\ref{comptestsec}.  This indicates that there are some differences
between the synthetic dataset we considered, and the observed data.
Much of the difference can be attributed to \sigmolmeas.  For the
synthetic datasets in Section \ref{testsec}, we simply considered a
uniform distribution in \sigmol, whereas the measured \sigmol\ in each
galaxy is not necessarily uniformly distributed.  Nevertheless, the
tests in Section \ref{testsec} showed that the Bayesian parameter
estimates are not sensitive to the intrinsic \sigmol\ distribution.
Further, note that the \sigscat\ estimates in Table \ref{bayesgals}
are all \aplt 0.1.  This suggests that a power law relationship can
reasonably describe the observed trends.  Taken together with the
results of the tests, any discrepancy in the true and prior
distributions of \sigmol\ and \sigscat\ likely does not strongly
effect the parameter estimates.

However, the assumed values of the conversion factors may play a
stronger role in the determination of the KS parameters.  We have
considered \sigmolmeas\ and \sigsfrmeas\ directly, such that there is
no uncertainty in the conversion from luminosities to those
quantities.  Here, we have assumed \XCO\ to be fixed at
2$\times10^{20}$ \Xunits, but in the next subsection we explore
variations in \XCO\ under the hierarchical Bayesian framework.

We can also compare the Bayesian result to the \OLSyx\ and \OLSxy\
estimates.  As described in Section \ref{comptestsec}, the \OLSyx\
should produce the most accurate estimates, because by definition it
estimates the mean value of \sigsfr\ given \sigmol.  To evaluate the
OLS slopes, we only need to consider the covariances and variances of
the data.  The ratio of Cov(\sigmolmeas, \sigsfrmeas)/Var(\sigmolmeas)
is 0.8, equivalent to the ratio of the ensemble in Group A, and by
definition to the \OLSyx\ fit slope of both the observed sample and
the synthetic data in Group A.  The \OLSyx\ slope is similar to the
Bayesian result, though the estimated uncertainty is very small, 0.04.
This uncertainty is solely statistical, does not account for
measurement errors, and its low value is primarily driven by the large
number of datapoints (981)\footnote{The uncertainty in the slope also
depends on the variance in $x$, i.e., on the width of the $x$
distribution.  If the $x$ distribution is broader, then the
uncertainty on the slope decreases. Because measurement errors broaden
the distribution of $x$, they also artificially decrease the
uncertainty in the slope estimate when they are ignored.}.
Consequently, the OLS methods significantly underestimate the errors,
and they do not provide the parameter estimates of the individuals
galaxies when all data are fit simultaneously.  We therefore caution
against the use of such statistical uncertainties for characterizing
the quality of the linear regression on large datasets.

To assess the difference between the hierarchical modeling and the
parameter estimates inferred from all data, we perform a direct
Bayesian fit to the ensemble.  For this analysis, we do not require a
group level in the hierarchical model.  In this case, we are only
interesting in estimating $N$, $A$, and the scatter term \sigscat\ for
the ensemble, shown in the last panel of Figure \ref{bayesgalfig}.  We
thus simplify the hierarchical model described in Section
\ref{hiermsec} by eliminating the group level, and employ uninformed
priors on the slope, intercept, and scatter terms.  We treat
measurement uncertainties in the same manner as described in Section
\ref{hiermsec}.  The posterior median of the Bayesian regression
analysis of the combined data yields $N$=0.90, with \tsig\ range
[0.88, 0.95].  This result is substantially different from the
parameter estimates from the full hierarchical model, and would lead
to different conclusions regarding the KS relationship.  Namely,
though the inferred slope is still less than unity, the slope
estimates for two of the individual galaxies, M51 and NGC 6946, are
excluded at the 95\% level.  Related to that discrepancy, the limited
range in the \tsig\ estimated slope of the ensemble suggests that
there may be a ``universal'' KS index.  However, the full hierarchical
Bayesian result provides a larger range for the estimated group
parameters and unambiguous differences between individual galaxy
parameters.  That indicates that there is likely no universal KS
relationship which can accurately describe the \sigsfr - \sigmol\
relationship for all galaxies, if our assumptions about the conversion
factors are valid, as further discussed in the next section.  The
variations in KS relationships between galaxies is clearly apparent
simply by inspecting the datapoints in Figure \ref{bayesgalfig},
affirming the variable star formation behaviour in different galaxies
identified by the hierarchical Bayesian fit.\footnote{If the KS
relationship were ``universal'', then the population variances of the
intercept and slope, $v_{\rm A}$ and $v_{N}$ respectively, would be
(very close) to zero.  That we find non-zero values $v_{\rm A}=0.09$
and $v_{N}=0.05$ also imply that there is no universal KS relationship
under this framework.}

\subsection{Variations in the \XCO\ factor} \label{Xsec}

In the analysis thus far, we have assumed that the conversion from
observed intensities to the star formation rates and molecular gas
surface densities are exact.  We thus implicitly assume that \sigsfr\
and \sigmol\ are directly measured.  But in actuality, the observed
quantities are UV, 24 \micron, and CO intensities, which are converted
to star formation rates and gas surface densities using constant
conversion factors.  Uncertainties in the conversion factors have thus
not been considered in the regression analysis.

A major advantage of the hierarchical framework is that uncertainties
in any quantity can be easily incorporated - for instance by including
additional levels for the conversion factors.  By explicitly
implementing another level for the conversion factors, the
hierarchical analysis can probabilistically ascertain whether certain
values of the conversion factors drive tighter KS relationships. Here,
we demonstrate the capability of the method to treat uncertainties in
the \XCO, allowing for independent variations in \XCO\ between each
datapoint.  In this framework, we are simply considering statistical
variations in \XCO.  As we describe below, observational and
theoretical efforts may further constrain this conversion factor.

In using the observed CO intensity \ICOmeas\ rather than \sigmolmeas,
we assume a measurement uncertainty similar to that described in
Section \ref{uncsec}.  Accordingly, we consider a linear regression of
the form:
\begin{equation}
\log(\Sigma_{\rm SFR}) = A + N \log (X_{\rm CO} I_{\rm CO}) +
\epsilon_{\rm scat}
\label{KSlaw_X}
\end{equation}
In this case, we need another level for the distribution of \XCO:
\begin{equation}
X_{\rm CO} \sim \mathcal{U}({\rm min} X_{\rm CO}, {\rm max} X_{\rm CO}),
\label{Xprior}
\end{equation}
where $\mathcal{U}$ is a uniform distribution between min\XCO\ and
max\XCO.  We have explored different extents for the ranges in \XCO,
including up to an order of magnitude, and find insignificant
differences in the results, either on the individual or group level.
The \tsig\ ranges increases slightly, but the mean/median of the
posterior does not change.  This indicates that independent variations
in \XCO\ do not have a strong influence on the KS parameters for this
sample.

As an example, Table \ref{bayesgalsXvar} shows the individual and
group slopes and scatter terms from the sample where \XCO\ is allowed
to vary from min\XCO=0.5$\times10^{20}$ and max\XCO=8$\times10^{20}$
\Xunits, allowing for over an order of magnitude variation in \XCO.
The median of the group slope is 0.79, and the \tsig\ interval is 0.58
to 1.0, which is slightly larger than the range shown in Table
\ref{bayesgals}.  There are hardly any differences between these
results and the parameter estimates when \XCO\ is fixed.\footnote{The
  model with fixed \XCO\ can be considered a special case of the more
  general model allowing for \XCO\ variations, but with a delta
  function on 2$\times10^{20}$ \Xunits\ as the prior.}  One difference
is that the slopes are systematically lower by \aplt 0.1.  Yet, there
is significant overlap between the results, considering the full 95\%
intervals, so there is no strong evidence of a decrease in slopes.
Further, the \sigscat\ values are similar between the two fits, again
suggestive that varying \XCO\ (by about an order of magnitude) does
not produce tighter KS relationships.  Similarly, the hierarchical
Bayesian fit does not provide any evidence for a preferred value of
\XCO.

\begin{table}
 \centering
 \begin{minipage}{90mm}
  \caption{Bayesian estimated slopes and scatter for the seven spiral
    galaxies in B08, where the \XCO\ factor is allowed to vary from
    0.5$\times10^{20}$ to 8$\times10^{20}$ \Xunits}
  \begin{tabular}{cccc}
  \hline
  \hline
 Subject & Bayes $N$ & Bayes $2\sigma_N$ & Bayes \sigscat  \\
\hline
NGC 5194 (M51) & 0.67 & [0.57, 0.78] & 0.07 \\
NGC 5055   & 0.86 & [0.78, 0.94] & 0.04 \\
NGC 3521   & 0.88 & [0.76, 1.02] & 0.05 \\
NGC 6946   & 0.74 & [0.66, 0.81] & 0.13 \\
NGC 628    & 0.65 & [0.45, 0.85] & 0.05 \\
NGC 3184   & 0.84 & [0.72, 0.97] & 0.05 \\
NGC 4736   & 0.86 & [0.59, 1.10] & 0.09 \\
\hline
{\bf Group Parameters}  & {\bf 0.79} & {\bf [0.58, 1.0]} & 0.14 \\
\hline
\end{tabular}
\label{bayesgalsXvar}
\end{minipage}
\end{table}

For this initial investigation into the effect of uncertainty in the
conversion factor, we assume that \XCO\ does not correlate with any
other property.  This preliminary analysis has only allowed for a
modest variation in \XCO, with no constraint for each datapoint beyond
the range set by conditional relationship (\ref{Xprior}).  Recent
theoretical efforts have indicated that \XCO\ indeed varies with other
physical properties, such as metallicity \citep{Glover&Maclow11,
Shetty+11a}, gas temperature, velocity dispersion
\citep{Narayanan+11a, Shetty+11b} or with molecular gas content
\citep{Feldmann+12b, Narayanan+12}.  Observational investigations are
also finding evidence for variable \XCO\ in individual galaxies
\citep[e.g.][]{Sandstrom+12}.  Such systematic variations would result
in different KS parameters \citep[see, e.g.][]{Ostriker&Shetty11,
Narayanan+12}.  We will investigate how the KS relationship is
affected by these systematic variations in \XCO\ in future work.  To
improve the parameter estimates, it would be helpful to increase the
number of sample galaxies and to incorporate uncertainties in
converting IR and UV luminosities to \sigsfr.  We can further explore
other parametrizations in \XCO, e.g. as a function of $I_{\rm CO}$,
and either for individual galaxies or for the population.  Such
modelling will allow for a study into the correlation between the
parameters, e.g. \XCO\ with \sigmol, in addition to fully treating
measurement and conversion factor uncertainties.

\section{Discussion \& Summary} \label{summarysec}

\subsection{Advantages of hierarchical Bayesian modelling}

We have demonstrated that the Bayesian method described in Section
\ref{methosec} can accurately recover the regression parameters from a
sample containing hierarchical structure.  Hierarchical data refers to
any set with numerous measurements from individuals within a group
(e.g. measurements from a sample of galaxies, with a number of
measurements per galaxy).  The method self-consistently treats
measurement uncertainties, so that the resulting parameter estimates
are PDFs which account for any uncertainty in the modelling process.

The Bayesian method is shown to be superior to the commonly employed
OLS estimates that usually do not consider the hierarchical structure
in the dataset.  Besides accounting for the measurement uncertainties,
the hierarchical nature of the dataset is preserved, as the Bayesian
method simultaneously estimates the parameters of both the individuals
as well as the population.  Under this framework, differences between
individuals are clearly exposed, and the range in plausible group
parameters self-consistently permits for variations between
individuals.

Another key aspect of the hierarchical Bayesian method is the robust
estimation of the scatter \sigscat\ about the fit regression line, for
both individuals subjects as well as the group.  When assessing the KS
relationship from observations, there is a trend of decreasing
\sigscat\ with increasing aperture size \citep[e.g.][]{Thilker+07,
  Kennicutt+07, Rahman+11}.  Theoretical efforts have considered
\sigscat\ and its relationship to sampling size.  It may be due to
variable sampling of the underlying cloud mass spectrum
\citep{Calzetti+12}, influence on averaging timescale, metallicity
\citep{Dib11}, background radiation field \citep{Feldmann+11}, or
simply transient effects \citep[e.g.][]{Shetty&Ostriker08}.  With
credible knowledge of the measurement uncertainties (e.g. $\sigma_{\rm
  mol}$ and $\sigma_{\rm SFR}$), the Bayesian method will provide
accurate estimates of the intrinsic scatter in the KS relationship.
Future work considering the possibility of variable conversion factors
for deriving \sigsfr\ and \sigmol\ will provide even more accurate
estimates of the intrinsic scatter, as the scatter due to the
systematics can be robustly separated from the intrinsic scatter.
These estimates of the intrinsic scatter can thereafter be used for
quantitative comparisons with theoretical results.

Recently, new statistical methods have been considered beyond the
bisector to estimate the KS parameters.  Namely, \citet{Blanc+09}
create numerous \sigsfr$-$\sigmol\ grids, considering different model
KS parameters, a scatter term, and different realizations of noise.
Then, they assess the \chisq\ between all the model grids and that
produced from observations of M51.  The distribution of \chisq\ values
identifies the best-fit parameters and their uncertainties.  As this
Monte Carlo approach considers the scatter term as well as treats
uncertainties, it should be more accurate than the typical OLS
bisector.  It has also recently been utilized in Leroy et al. (2012,
Submitted) on the HERACLES sample.  It will be interesting to
ascertain whether such a method can accurately recover the individual
and group parameters from a sample of galaxies, and how it compares to
the hierarchical Bayesian method presented here.

\subsection{The KS relationship in galaxies} \label{interpsec}

As discussed in the introduction, the KS law is a widely investigated
relationship between the star formation rate and molecular gas surface
density.  The initial works using galaxy wide averages estimated
N$\approx$1.5.  Follow up work using background subtracted emission
from resolved observations by \citet{Kennicutt+07} and \citet{Liu+11}
recovered this result.  On the other hand, kpc-scale observations of a
large sample of galaxies have inferred slopes closer to unity
(e.g. B08, \citealt{Bigiel+11, Leroy+08}, Leroy et al. 2012 Submitted,
\citealt{Rahman+12}).

After verifying the accuracy of the hierarchical Bayesian method,
which models how the mean value of log(\sigsfr) depends on
log(\sigmol), we applied the method to estimate the parameters of the
KS relationship in the sample of disk galaxies from B08.  The fitting
results from both the test samples (Section \ref{testsec}) and the B08
sample suggest that previous methods have overestimated the KS slopes.
The hierarchical Bayesian method produces slopes which are slightly
lower than unity, especially when considering the full 95\% confidence
interval for some of the individual galaxies.

Before interpreting our results from this analysis, we note a number
of important caveats which should be kept in mind.  First, as we have
discussed throughout this work, we have directly considered
\sigsfrmeas\ and \sigmolmeas.  Yet, these are not directly measured.
The star formation rates and surface densities are estimated from the
observables (e.g., CO, UV and IR intensities) using constant
conversion factors.  In the measurement model
(Eqns. \ref{measmod1}-\ref{measmod2}), we have only accounted for
random statistical errors associated with observational noise.  Yet,
systematic errors may arise in the conversion between intensities to
\sigsfrmeas\ and \sigmolmeas.  Under a hierarchical framework,
uncertainties in these conversion factors can be naturally handled,
including correlated errors between the various coefficients.  In
future work, we will self-consistently treat the uncertainties in
these conversion factors for a larger sample of galaxies.  This may
reveal interesting new aspects of the star formation and gas
properties in the ISM.

Additionally, we have not considered the different approaches to
estimating star formation rates on $\sim$kpc scales in galaxies.  As
\citet{Kennicutt+07}, \citet{Liu+11}, and \citet{Rahman+11}
demonstrate, subtracting off a diffuse component can have significant
impact on the scaling relationships: it may increase the KS slope,
because the relative contribution of diffuse emission is largest where
\sigsfr\ is small (see also \citealt{Leroy+12}).  A similar argument
could possibly be made regarding diffuse CO emission, which may not be
directly related to recent star formation on $\sim$kpc scales
\citep{Rahman+11}.  B08 treat some of these aspects differently from
\citet{Kennicutt+07} and \citet{Liu+11} and we refer the reader to
these studies for further details.  We note that \citet{Leroy+12}
explore some of these effects, in particular the contribution from an
evolved stellar population to the IR star formation tracer, and
suggest that the uncertainty due to diffuse emission in the estimated
index $N$ is $\sim0.15\times N$.  This may push some of the indices of
individual galaxies, such as NGC 3184 and NGC 4736 to values greater
than unity.  But, for the galaxies with the lowest indices, such as
M51 and NGC 6946, the data would still favor a sub-linear KS
relationship.

One of the results from the hierarchical Bayesian fit on the B08
sample is that there are significant galaxy-to-galaxy variations in
the slopes and intercepts, as already noted by B08.  The comparison of
the PDFs of parameter estimates between certain galaxies indicates
that to a very high degree of confidence galaxies have different KS
parameters.  For instance, at $>$ 95\% M51 and NGC 3184 have different
intercepts.  Similarly, the slopes are also substantially different,
as the posterior median value for NGC 3184, 0.92, is ruled out for
M51, though there is a very low probability that both galaxies have
slopes between 0.79 and 0.83.

The significant galaxy-to-galaxy variation leads to a group
distribution of the slope with a large dispersion, $\sim$ 0.4.  This
wide distribution suggests that there is likely no single, or
``universal'' KS law that is applicable over all disk galaxies.  This
result supports the idea that other physical properties, such as
metallicity, gas fraction, stellar mass, turbulence levels, magnetic
fields, or galaxy environment, among other factors, affect the star
forming properties of a galaxy besides the gas surface density.
Recent observational results support this description
(e.g. \citealt{Leroy+09, Shi+11, Saintonge11b, Sandstrom+12}, Leroy et
al. 2012 Submitted).  Theoretical efforts are also considering the
impact of other physical processes besides just the gas surface
density \citep[e.g.][]{Stinson+06, Ostriker+10, Kim+11,
  Ostriker&Shetty11, Krumholz+12, Shetty&Ostriker12, Glover&Clark12,
  Federrath&Klessen12}.

The results from the hierarchical Bayesian fit on the seven spiral
galaxies has produced slopes which are systematically lower than
previous analyses.  The posterior median values for all seven
individual galaxies are all less than unity.  In four of the seven
galaxies, NGC 5194, NGC 5055, NGC 6946, and NGC 628, the slope
estimates are well below unity, even at the \tsig\ level.  For M51
(NGC 5194), the posterior median of the slope is 0.72, with values
\apgt 0.83 ruled out with 95\% confidence.\footnote{In fact, for M51,
the maximum slope in the posterior is 0.97, ruling out a linear slope
under the hierarchical Bayesian framework.}  The sub-linear slope in
M51 was even estimated by B08 with the bisector.  Moreover, using
\Halpha\ observations, and a Monte Carlo analysis of the KS
parameters, \citet{Blanc+09} estimated $N = 0.82 \pm 0.05$.  Taken
together, these results strongly point to a sub-linear KS slope for
M51.

The Bayesian inferred group slope is 0.84, with \tsig\ range [0.63,
1.0].  This range is consistent with the variable estimated slopes
from the individual galaxies.  Future work on a larger sample is
needed to confirm this \tsig\ result.  One interpretation of a
sub-linear relationship is that the gas surface density estimated from
the detected CO emission is not all associated with star formation.
The plausible interpretation is that at higher CO luminosities, the
relative fraction of diffuse molecular gas that is not associated with
star formation increases.  The KS index may provide a quantitative
measure of this fraction.

A sub-linear slope also suggests that the efficiency of star formation
(here defined as the star formation rate normalized to H$_2$ mass)
decreases for increasing gas surface density.  Equivalently, the
computed gas depletion time is not constant, and increases with
increasing gas surface density.  Of course, this depletion time refers
to all the gas that is {\it detected}, and if more non-star forming
gas is contributing to the observed emission, then the depletion time
calculation may include superfluous gas not associated with star
formation.  This is consistent with the picture that at high CO
intensities, relatively more diffuse gas is contributing to the
emission compared to lower intensity regions, where CO is only tracing
dense gas.  Yet, star formation only occurs in the dense most
(gravitationally-bound) regions of the ISM, regardless of whether CO
is present in dense or diffuse gas.

We have also applied a Bayesian regression fit to the publicly
available star formation rates and gas surface densities in M51
inferred by \citet{Kennicutt+07}.  In their investigation,
\citet{Kennicutt+07} measure \sigsfrmeas\ by subtracting a background
from the 24 \micron\ measurements, assuming that some fraction of the
dust is heated by an older stellar population and thus is not related
to recent star formation.  From our analysis, we estimate slopes (at
95\% confidence) in the range [1.25, 1.59] for the 13\arcsec\ (520 pc)
data, and [0.80, 1.22] from the 45\arcsec\ (1850 pc) data.  These
slopes are larger than the range obtained from the analysis on the B08
sample, as well as the M51 study by \citet{Blanc+09}.  This
discrepancy is possibly due to methodological differences, e.g. a
diffuse, non-star formation related component in the IR emission (see
discussion above), a topic that is currently extensively under debate
\citep{Liu+11, Rahman+12, Leroy+12}.  Nevertheless, the Bayesian
results are rather different from the KS parameters estimated in
\citet{Kennicutt+07}, in that there is a significant difference
between the 13\arcsec\ and
45\arcsec\ data.\footnote{\citet{Kennicutt+07} employ a bi-linear
  (FITEXY) fitting routine, and as discussed by \citet{Calzetti+12},
  the inferred index can be dependent on the fitting method, as well
  as the size of the sampled regions \citep[see also][]{Rahman+11}.}
Namely, the KS index decreases with increasing beam size.  A decrease
in slope with beam size would be consistent with the interpretation
offered above for the sub-linear slopes in the B08 sample.  In the CO
bright ISM, if there is a contribution from diffuse or dense gas not
associated with star formation, then increasing the sampling area
might also increase the contribution from this component to \sigmol,
without a corresponding increase in \sigsfr.  This scenario would lead
to a systematic decrease of the KS index with increasing beam sizes.

Supporting this description, \citet{Elmegreen93} postulated the
presence of diffuse molecular clouds.  In the ISM where the pressure
is sufficiently high, H$_2$ molecules can form in regions which are
not necessarily self-gravitating.  \citet{Elmegreen93} suggests that
due to local and transient variations in the ISM, a high pressure
region may form molecules.  Similarly, the diffuse molecular gas may
be returned to the atomic phase before it becomes self-gravitating due
to a decrease in pressure or an increase in UV radiation (e.g. from
nearby star formation).  Thus, at any given time there may be some
fraction of the ISM which is in molecular form but is returned to the
atomic phase before star formation occurs in that diffuse molecular
component.

Whether the sub-linearity in the KS index is a sign of such non-star
forming molecular gas remains to be tested.  One important
consideration is whether CO can faithfully trace this diffuse
molecular component.  \citet{Glover&MacLow07II} showed that due to
effective self-shielding, H$_2$ can exist in diffuse regions, whereas
CO is easily photodissociated.  In these regions, CO emission is not a
reliable tracer of molecular gas.  As a result \XCO\ can vary
depending on environment \citep{Glover&Maclow11,
  Shetty+11a,Shetty+11b}.  Yet, in the denser regions of a Galaxy,
e.g. towards the centre, the ISM pressure and density may be
sufficiently high, allowing CO to form alongside H$_2$ in clouds which
are not self-gravitating nor star forming, thereby resulting in the
sub-linear KS relationship we find here.

Our analysis has focused on the relationship between star formation
and CO traced gas on approximately kpc-scales in external galaxies.
Accordingly, the interpretation that a sub-linear KS relationship is a
sign of molecular gas unassociated with star formation can only be
applicable on those large scales.  Detailed analysis on smaller scales
should also be conducted to verify the presence of such gas.  Indeed,
the recent observational analysis by \citet{Longmore+12b} has
indicated that the relationship between the star formation rate and
gas density in the Central Molecular Zone (CMZ) within \aplt500 pc
from the Galactic Centre is discrepant from that inferred in the main
disc.  Namely, it appears that the star formation rate is about an
order of magnitude lower than expected given the high gas densities in
the CMZ, compared to a linear KS relationship inferred from previous
investigations.  The CMZ environment is rather different than the main
disc of the Milky Way, in that the mean density and molecular
fractions are significantly higher \citep{Bally+87, Morris&Serabyn96},
as well as the turbulent velocities \citep{Bally+87,
  Miyazaki&Tsuboi00, Oka+98, Oka+01, Shetty+12}.  As suggested by
\citet{Longmore+12b}, these environmental differences may contribute
to the discrepant star formation behavior.  Such variable ISM
environments, including the higher fraction of molecular gas towards
galactic centers, may also be extant in external galaxies, and
contribute to the sub-linear KS relationship we find here.

The sub-linear KS relationship we estimate for some individual
galaxies is only applicable in the ISM where \sigmol \aplt\ 100
\msunpc.  As originally discussed by B08, there is evidence for a
steeping in the KS index at higher surface densities.
\citet{Ostriker&Shetty11} and \citet{Shetty&Ostriker12} postulated a
KS relationship with $N\approx$2 for the molecular dominanted ISM of
starbursts if supernovae driven turbulent pressure balances the
vertical weight of the disk, leading to the self-regulation of star
formation. \citet{Ostriker&Shetty11} showed that if the \XCO\ factor
various continuously with surface density, then the best fit KS
relationship to an observed sample of starbursts by \citet{Genzel+10}
has $N\approx 2$.  \citet{Narayanan+12} confirmed this prediction,
using \XCO $\propto I_{\rm CO}^{-0.3}$, a result based on numerical
modeling of a suite of galaxy simulations.  The apparent break in the
KS relationship between the starbursts, with super-linear slopes, and
more quiscent ISM, for which this work has found $N < 1$ for some
individual galaxies, may be indicative of fundamemtal differences in
the properties of the ISM between the regimes.  One possibility is
that the relative amount of dense and diffuse molecular gas may differ
between these regimes.

These ideas will have to be further scrutinized quantitatively, both
theoretically and observationally.  Recent theoretical models are now
capable of tracking the formation of H$_2$ and CO in ISM simulations
\citep[e.g.][]{Glover&MacLow07I,Glover&MacLow07II, Glover+10}.
Simulations of colliding flows with time-dependent chemistry show that
CO forms in clouds more rapidly and more pervasively than the
formation of stars \citep{Clark+12}.  Extensions of that work are
showing that the star formation rate does not increase as rapidly with
CO abundance in the most massive clouds, compared to the trend at
lower masses and lower CO densities (P. Clark et al. In Prep).  This
scenario is consistent with a sub-linear KS index due to the presence
of CO in gas which is not later converted into stars.

An extension of the hierarchical Bayesian analysis performed here,
including a larger sample and varying sampling beams, may be needed to
confirm these trends.  Hierarchically assessing the KS parameters may
quantify the fraction of molecular gas not associated with star
formation in individual galaxies.  Given the variation in KS
relationships in individual galaxies, we expect to find differences in
the diffuse molecular gas fraction from galaxy to galaxy.  Explicitly
accounting for uncertainties in estimating \sigsfr\ and \sigmol\ may
also reveal the correlations between various conversion factors, such
as \XCO\ and the conversion from IR luminosity to \sigsfr.  The
hierarchical Bayesian approach is ideally suited for these analyses.
In future work, we will deploy the hierarchical model on larger
datasets, which should further advance our understanding of star
formation in the ISM.

\subsection{Summary of results}

We have introduced a Bayesian linear regression method for inferring
the parameters of the KS law, which consistently treats measurement
uncertainties, as well as the hierarchical structure of datasets.
After demonstrating the accuracy of the method on synthetic datasets,
we applied the method to estimate the KS parameters from observations
of disk galaxies by B08.

Our main results are as follows:

1.  A hierarchical Bayesian method is well suited for linear
regression of structured data with measurement uncertainties,
e.g. numerous measurements of individuals within a group.  The method
simultaneously fits the regression parameters of each individual and
the group, and provides PDFs of the slope, intercept and scatter
terms.  We demonstrate that the posterior, which includes well defined
uncertainty estimates, accurately recovers the parameters of synthetic
hierarchical data (Sections \ref{methosec} and \ref{testsec}).

2. We compared the Bayesian result with the OLS ($y|x$), ($x|y$), and
bisector fits on the synthetic datasets.  From the test on synthetic
datasets, the Bayesian method provides the most accurate estimates of
underlying parameters for both individuals and the population.  The
commonly employed bisector usually overestimates the slopes and
underestimates the intercept.  We discuss that the reason for the
discrepancies is that OLS methods provide different summaries of the
joint distribution.  We therefore recommend against the use of the
bisector for linear regression of data with measurement uncertainties
(Section \ref{testsec}).

3. When applied to observed data of spiral galaxies in B08 to estimate
the KS parameters, we obtained slopes that are lower than previous
results.  In four of the seven galaxies, NGC 5194, NGC 5055, NGC 6946,
and NGC 628, the slope estimates are well below unity, even at the
\tsig\ level.  For NGC 5194, the posterior median of the slope is
0.72.  For the group slope and intercept, the posterior median is 0.84
and $-$3.00, with \tsig\ range [0.63, 1.0] and [$-3.3, -2.7$],
respectively.  The posterior median of the intrinsic scatter, assuming
25\% and 50\% uncertainty in \sigmolmeas\ and \sigsfrmeas,
respectively, is 0.14 dex.  Previous bisector results on the ensemble
overestimated the slopes, as confirmed by the large discrepancy in the
\OLSyx\ and \OLSxy\ estimates.  We also noted that a direct Bayesian
linear regression on the ensemble provides limited ranges in the KS
parameters, thereby attesting to the necessity for treating the
hierarchical nature of datasets in order to clearly identify the
differences between individual galaxies (Section \ref{ressec}).

4. A sub-linear KS relationship or a decreasing KS index with
increasing beam size for some individual galaxies may be indicative of
molecular gas which is not forming stars.  At low densities or
metallicities, CO bright regions may be directly associated with star
formation.  But at increasingly higher CO luminosities, diffuse or
dense molecular gas not associated with star formation may be
contributing to the observed emission.  This situation would
correspondingly result in a sub-linear KS relationship, so that the
gas depletion time is not constant but rather increases with CO traced
\sigmol.  As we find significant variation from galaxy-to-galaxy, this
scenario may be applicable to some galaxies, e.g. those with high
molecular gas fractions, but not others (Section \ref{interpsec}).

To improve our analysis, in future work we will consider a larger
sample, include a treatment of uncertainties in the conversion factors
and star formation rate calibrations (effects of diffuse emission not
related to star formation), and explicitly consider the effects of
other physical properties of the source.  As the hierarchical Bayesian
framework presented here is well-suited for treating any source of
uncertainty, it can naturally handle variations in the conversion
factors, and may provide additional insights into the properties of
star formation in the ISM.

\section*{Acknowledgements}

We are especially grateful to Paul Clark, Bruce Elmegreen, Simon
Glover, Ralf Klessen, Steven Longmore, Eve Ostriker, and Greg Stinson
for extensive discussions on star formation in molecular gas as well
as comments on the draft.  We also thank Richard Allison, Elly
Berkhuijsen, Chris Hayward, Lukas Konstandin, Adam Leroy, Amelia
Stutz, and Benjamin Weiner for valuable input on statistical
inference, IR emission, and star formation.  We appreciate
constructive comments from an anonymous referee that improved this
work.  RS acknowledges support from the Deutsche
Forschungsgemeinschaft (DFG) via the SFB 881 (B1 and B2) ``The Milky
Way System,'' and the SPP (priority program) 1573.

\begin{sidewaystable*}[h]
\vspace{-10cm}
{\bf Table 3.} OLS$^1$ estimated parameters for Test Group A \\
\centering                            
  \begin{tabular}{ccccccccc}
  \hline
  \hline
 Subject & True $A$ & True $N$ & \OLSyx\ $A$ & \OLSyx\ $N$ & \OLSxy\ $A$ & \OLSxy\ $N$ & Bisector $A$ & Bisector $N$  \\
\hline
Test Galaxy A1  & $-$2.77 & 0.72 & $-$2.75 \ppm\ 0.11 & 0.71 \ppm\ 0.08  & $-$2.89 \ppm\ 0.29 & 0.83 \ppm\ 0.15 & $-$2.82 & 0.77  \\
Test Galaxy A2  & $-$3.21 & 0.88 & $-$3.23 \ppm\ 0.13 & 0.86 \ppm\ 0.10  & $-$3.39 \ppm\ 0.27 & 0.99 \ppm\ 0.12 & $-$3.31 & 0.92  \\
Test Galaxy A3  & $-$3.18 & 0.89 & $-$3.13 \ppm\ 0.14 & 0.87 \ppm\ 0.11  & $-$3.31 \ppm\ 0.26 & 1.03 \ppm\ 0.12 & $-$3.22 & 0.95  \\
Test Galaxy A4  & $-$2.81 & 0.78 & $-$2.88 \ppm\ 0.13 & 0.79 \ppm\ 0.10  & $-$3.06 \ppm\ 0.27 & 0.95 \ppm\ 0.14 & $-$2.96 & 0.87  \\
Test Galaxy A5  & $-$2.87 & 0.74 & $-$2.88 \ppm\ 0.13 & 0.73 \ppm\ 0.10  & $-$3.07 \ppm\ 0.32 & 0.90 \ppm\ 0.15 & $-$2.97 & 0.81  \\
Test Galaxy A6  & $-$3.22 & 0.91 & $-$3.12 \ppm\ 0.13 & 0.86 \ppm\ 0.10  & $-$3.28 \ppm\ 0.24 & 1.00 \ppm\ 0.11 & $-$3.20 & 0.93 \\
Test Galaxy A7  & $-$2.82 & 0.92 & $-$2.75 \ppm\ 0.14 & 0.87 \ppm\ 0.11  & $-$2.92 \ppm\ 0.22 & 1.03 \ppm\ 0.12 & $-$2.92 & 0.95  \\
\hline
{\bf Group Parameters}  & {\bf$-$2.98} & {\bf 0.83} & {\bf $-$2.96 \ppm\ 0.06} & {\bf 0.81 \ppm\ 0.05} & {\bf $-$3.22 \ppm\ 0.11} & {\bf 1.04 \ppm\ 0.05} & {\bf $-$3.1} & {\bf 0.92} \\
\hline
\footnotetext[0]{$^1$ \tsig\ uncertainties are provided for \OLSyx\ and \OLSxy\ estimated parameter.}
\end{tabular}
\label{OLSgrpA}
\end{sidewaystable*}

\vspace{15cm}
\begin{sidewaystable*}
\vspace{15cm}
{\bf Table 4.} OLS$^1$ estimated parameters for Test Group B \\
\centering                            
  \begin{tabular}{ccccccccc}
  \hline
  \hline
 Subject & True $A$ & True $N$ & \OLSyx\ $A$ & \OLSyx\ $N$ & \OLSxy\ $A$ & \OLSxy\ $N$ & Bisector $A$ & Bisector $N$  \\
\hline
Test Galaxy B1 & $-$2.29 & 0.84 & $-$2.26 \ppm\ 0.12 & 0.82 \ppm\ 0.09 & $-$2.40 \ppm\ 0.17 & 0.93 \ppm\ 0.12 & $-$2.33 & 0.88  \\
Test Galaxy B2 & $-$2.53 & 0.92 & $-$2.55 \ppm\ 0.13 & 0.90 \ppm\ 0.10 & $-$2.70 \ppm\ 0.18 & 1.03 \ppm\ 0.11 & $-$2.62 & 0.96  \\
Test Galaxy B3 & $-$2.15 & 0.96 & $-$2.46 \ppm\ 0.14 & 0.94 \ppm\ 0.11 & $-$2.64 \ppm\ 0.16 & 1.09 \ppm\ 0.11 & $-$2.55 & 1.02  \\
Test Galaxy B4 & $-$2.26 & 0.92 & $-$2.32 \ppm\ 0.14 & 0.93 \ppm\ 0.10 & $-$2.47 \ppm\ 0.15 & 1.06 \ppm\ 0.10 & $-$2.39 & 0.99  \\
Test Galaxy B5 & $-$2.33 & 1.00 & $-$2.33 \ppm\ 0.13 & 0.99 \ppm\ 0.10 & $-$2.48 \ppm\ 0.13 & 1.12 \ppm\ 0.09 & $-$2.40 & 1.05  \\
Test Galaxy B6 & $-$2.54 & 1.12 & $-$2.44 \ppm\ 0.13 & 1.07 \ppm\ 0.10 & $-$2.57 \ppm\ 0.11 & 1.18 \ppm\ 0.08 & $-$2.51 & 1.12  \\
Test Galaxy B7 & $-$2.12 & 0.95 & $-$2.03 \ppm\ 0.14 & 0.90 \ppm\ 0.10 & $-$2.21 \ppm\ 0.13 & 1.05 \ppm\ 0.11 & $-$2.12 & 0.98  \\
\hline
{\bf Group Parameters}  & {\bf$-$2.37} & {\bf 0.96} & {\bf $-$2.34 \ppm\ 0.06} & {\bf 0.93 \ppm\ 0.05} & {\bf $-$2.57 \ppm\ 0.06} & {\bf 1.13 \ppm\ 0.04} & {\bf $-$2.45} & {\bf 1.03} \\
\hline
\footnotetext[0]{$^1$ \tsig\ uncertainties are provided for \OLSyx\ and \OLSxy\ estimated parameters.}
\end{tabular}
\label{OLSgrpB}
\end{sidewaystable*}

\vspace{-15cm}
\begin{sidewaystable*}
\vspace{-15cm}
{\bf Table 5.} Adopted and Bayesian inferred parameters for Test Group C \\
\centering                            
  \begin{tabular}{cccccccccc}
  \hline
  \hline
 Subject & Datapoints & True $A$ & True $N$ & True Scatter$^1$ & Bayes $A$ & Bayes $2\sigma_A$ & Bayes $N$ & Bayes $2\sigma_N$ & Bayes \sigscat  \\
\hline
Test Galaxy C1  & 5  & $-$3.00 & 1.50 & 0.3 & $-$2.63 & [$-$3.00, $-$2.33] & 1.22 & [1.00, 1.49] & 0.17 \\
Test Galaxy C2  & 7  & $-$2.95 & 1.46 & 0.3 & $-$2.58 & [$-$2.90, $-$2.30] & 1.22 & [1.01, 1.47] & 0.17 \\
Test Galaxy C3  & 9  & $-$2.89 & 1.42 & 0.3 & $-$2.60 & [$-$2.93, $-$2.34] & 1.21 & [1.00, 1.46] & 0.17 \\
Test Galaxy C4  & 11 & $-$2.84 & 1.37 & 0.3 & $-$2.63 & [$-$2.96, $-$2.37] & 1.19 & [0.99, 1.43] & 0.17 \\
Test Galaxy C5  & 13 & $-$2.79 & 1.33 & 0.3 & $-$2.58 & [$-$2.88, $-$2.33] & 1.20 & [1.00, 1.43] & 0.17 \\
Test Galaxy C6  & 15 & $-$2.74 & 1.29 & 0.3 & $-$2.54 & [$-$2.82, $-$2.30] & 1.12 & [0.93, 1.32] & 0.17 \\
Test Galaxy C7  & 17 & $-$2.68 & 1.25 & 0.3 & $-$2.55 & [$-$2.81, $-$2.31] & 1.17 & [0.98, 1.37] & 0.17 \\
Test Galaxy C8  & 19 & $-$2.63 & 1.21 & 0.3 & $-$2.58 & [$-$2.83, $-$2.34] & 1.15 & [0.97, 1.34] & 0.17 \\
Test Galaxy C9  & 21 & $-$2.58 & 1.16 & 0.3 & $-$2.59 & [$-$2.77, $-$2.30] & 1.14 & [0.97, 1.33] & 0.17 \\
Test Galaxy C10 & 23 & $-$2.53 & 1.12 & 0.3 & $-$2.48 & [$-$2.72, $-$2.26] & 1.07 & [0.90, 1.25] & 0.17 \\
Test Galaxy C11 & 25 & $-$2.47 & 1.08 & 0.3 & $-$2.45 & [$-$2.69, $-$2.24] & 1.10 & [0.93, 1.29] & 0.17 \\
Test Galaxy C12 & 27 & $-$2.42 & 1.04 & 0.3 & $-$2.43 & [$-$2.66, $-$2.22] & 1.07 & [0.91, 1.24] & 0.17 \\
Test Galaxy C13 & 29 & $-$2.37 & 0.99 & 0.3 & $-$2.41 & [$-$2.63, $-$2.20] & 1.04 & [0.88, 1.21] & 0.17 \\
Test Galaxy C14 & 31 & $-$2.32 & 0.95 & 0.3 & $-$2.32 & [$-$2.54, $-$2.13] & 0.99 & [0.83, 1.15] & 0.17 \\
Test Galaxy C15 & 33 & $-$2.26 & 0.91 & 0.3 & $-$2.31 & [$-$2.51, $-$2.11] & 0.93 & [0.78, 1.09] & 0.17 \\
Test Galaxy C16 & 35 & $-$2.21 & 0.87 & 0.3 & $-$2.31 & [$-$2.54, $-$2.11] & 0.94 & [0.78, 1.12] & 0.17 \\
Test Galaxy C17 & 37 & $-$2.16 & 0.86 & 0.3 & $-$2.21 & [$-$2.42, $-$2.02] & 0.89 & [0.74, 1.04] & 0.17 \\
Test Galaxy C18 & 39 & $-$2.11 & 0.78 & 0.3 & $-$2.20 & [$-$2.40, $-$2.02] & 0.86 & [0.72, 1.01] & 0.17 \\
Test Galaxy C19 & 41 & $-$2.05 & 0.74 & 0.3 & $-$2.18 & [$-$2.37, $-$2.00] & 0.83 & [0.68, 0.98] & 0.17 \\
Test Galaxy C20 & 43 & $-$2.00 & 0.70 & 0.3 & $-$2.14 & [$-$2.33, $-$1.96] & 0.79 & [0.65, 0.95] & 0.17 \\
\hline
{\bf Group C Parameters}  & {\bf 480} & {\bf$-$2.5} & {\bf 1.1} & {\bf
  0.3} & {\bf $-$2.43} & {\bf [$-$2.57, $-$2.31]} & {\bf 1.06} & {\bf
  [0.95, 1.17]} & {\bf 0.17} \\
\hline
\footnotetext[0]{$^1$ For this dataset, the intrinsic scatter is uniformly
  distributed.  Values refer to the width of the distribution,
  centered on 0.}
\end{tabular}
\label{BayesgrpC}
\end{sidewaystable*}

\vspace{15cm}
\begin{sidewaystable*}
\vspace{15cm}
{\bf Table 6.} OLS$^1$ estimated parameters for Test Group C \\
\centering                            
  \begin{tabular}{ccccccccccc}
  \hline
  \hline
 Subject & Datapoints & True $A$ & True $N$ & \OLSyx\ $A$ & \OLSyx\ $N$ & \OLSxy\ $A$ & \OLSxy\ $N$ & Bisector $A$ & Bisector $N$  \\
\hline
Test Galaxy C1  & 5  & $-$3.00 & 1.50 & $-$3.10 \ppm\ 0.13 & 1.53 \ppm\ 0.10 & $-$3.10 \ppm\ 0.07 & 1.54 \ppm\ 0.04 & $-$3.10 & 1.53  \\
Test Galaxy C2  & 7  & $-$2.95 & 1.46 & $-$2.83 \ppm\ 0.20 & 1.39 \ppm\ 0.15 & $-$2.85 \ppm\ 0.12 & 1.41 \ppm\ 0.08 & $-$2.85 & 1.40  \\
Test Galaxy C3  & 9  & $-$2.89 & 1.42 & $-$2.88 \ppm\ 0.16 & 1.41 \ppm\ 0.12 & $-$2.90 \ppm\ 0.10 & 1.43 \ppm\ 0.12 & $-$2.89 & 1.42  \\
Test Galaxy C4  & 11 & $-$2.84 & 1.37 & $-$2.84 \ppm\ 0.15 & 1.33 \ppm\ 0.11 & $-$2.87 \ppm\ 0.09 & 1.35 \ppm\ 0.05 & $-$2.86 & 1.34  \\
Test Galaxy C5  & 13 & $-$2.79 & 1.33 & $-$2.76 \ppm\ 0.12 & 1.34 \ppm\ 0.09 & $-$2.78 \ppm\ 0.08 & 1.35 \ppm\ 0.05 & $-$2.77 & 1.34  \\
Test Galaxy C6  & 15 & $-$2.74 & 1.29 & $-$2.60 \ppm\ 0.13 & 1.15 \ppm\ 0.10 & $-$2.62 \ppm\ 0.11 & 1.18 \ppm\ 0.07 & $-$2.61 & 1.16  \\
Test Galaxy C7  & 17 & $-$2.68 & 1.25 & $-$2.63 \ppm\ 0.17 & 1.22 \ppm\ 0.13 & $-$2.69 \ppm\ 0.12 & 1.27 \ppm\ 0.08 & $-$2.66 & 1.25  \\
Test Galaxy C8  & 19 & $-$2.63 & 1.21 & $-$2.69 \ppm\ 0.11 & 1.23 \ppm\ 0.09 & $-$2.72 \ppm\ 0.08 & 1.25 \ppm\ 0.06 & $-$2.71 & 1.24  \\
Test Galaxy C9  & 21 & $-$2.58 & 1.16 & $-$2.56 \ppm\ 0.10 & 1.17 \ppm\ 0.07 & $-$2.59 \ppm\ 0.07 & 1.19 \ppm\ 0.05 & $-$2.57 & 1.18  \\
Test Galaxy C10 & 23 & $-$2.53 & 1.12 & $-$2.48 \ppm\ 0.09 & 1.06 \ppm\ 0.07 & $-$2.51 \ppm\ 0.09 & 1.09 \ppm\ 0.06 & $-$2.50 & 1.08  \\
Test Galaxy C11 & 25 & $-$2.47 & 1.08 & $-$2.44 \ppm\ 0.11 & 1.09 \ppm\ 0.09 & $-$2.49 \ppm\ 0.10 & 1.13 \ppm\ 0.07 & $-$2.47 & 1.11  \\
Test Galaxy C12 & 27 & $-$2.42 & 1.04 & $-$2.41 \ppm\ 0.10 & 1.05 \ppm\ 0.08 & $-$2.45 \ppm\ 0.09 & 1.08 \ppm\ 0.07 & $-$2.43 & 1.07  \\
Test Galaxy C13 & 29 & $-$2.37 & 0.99 & $-$2.37 \ppm\ 0.11 & 1.01 \ppm\ 0.08 & $-$2.42 \ppm\ 0.11 & 1.05 \ppm\ 0.08 & $-$2.40 & 1.03  \\
Test Galaxy C14 & 31 & $-$2.32 & 0.95 & $-$2.25 \ppm\ 0.10 & 0.93 \ppm\ 0.08 & $-$2.31 \ppm\ 0.11 & 0.98 \ppm\ 0.08 & $-$2.28 & 0.95  \\
Test Galaxy C15 & 33 & $-$2.26 & 0.91 & $-$2.22 \ppm\ 0.08 & 0.86 \ppm\ 0.06 & $-$2.26 \ppm\ 0.10 & 0.89 \ppm\ 0.08 & $-$2.24 & 0.88  \\
Test Galaxy C16 & 35 & $-$2.21 & 0.87 & $-$2.23 \ppm\ 0.08 & 0.88 \ppm\ 0.07 & $-$2.28 \ppm\ 0.11 & 0.92 \ppm\ 0.08 & $-$2.26 & 0.90  \\
Test Galaxy C17 & 37 & $-$2.16 & 0.86 & $-$2.10 \ppm\ 0.09 & 0.80 \ppm\ 0.07 & $-$2.16 \ppm\ 0.12 & 0.85 \ppm\ 0.10 & $-$2.13 & 0.83  \\
Test Galaxy C18 & 39 & $-$2.11 & 0.78 & $-$2.09 \ppm\ 0.07 & 0.77 \ppm\ 0.05 & $-$2.12 \ppm\ 0.11 & 0.80 \ppm\ 0.08 & $-$2.10 & 0.78  \\
Test Galaxy C19 & 41 & $-$2.05 & 0.74 & $-$2.06 \ppm\ 0.07 & 0.74 \ppm\ 0.05 & $-$2.11 \ppm\ 0.12 & 0.78 \ppm\ 0.10 & $-$2.08 & 0.76  \\
Test Galaxy C20 & 43 & $-$2.00 & 0.70 & $-$2.01 \ppm\ 0.07 & 0.69 \ppm\ 0.05 & $-$2.06 \ppm\ 0.13 & 0.73 \ppm\ 0.10 & $-$2.04 & 0.71  \\
\hline
{\bf Group C Parameters}  & 480 & {\bf$-$2.5} & {\bf 1.1} & {\bf $-$2.34 \ppm\ 0.03} & {\bf 0.98 \ppm\ 0.03} & {\bf $-$2.44 \ppm\ 0.03} & {\bf 1.06 \ppm\ 0.03} & {\bf $-$2.39} & {\bf 1.02} \\
\hline
\footnotetext[0]{$^1$ \tsig\ uncertainties are provided for \OLSyx\ and \OLSxy\ estimated parameters.}
\end{tabular}
\label{OLSgrpC}
\end{sidewaystable*}

\bibliography{citations}
\bibliographystyle{mn2e}

\label{lastpage}

\end{document}